\documentclass[11pt,english]{article}
\usepackage{jheppub}
\usepackage{setspace}
\usepackage{bbm}

\newcommand{\nn}{\nonumber}

\makeatother

\begin{document}
\preprint{MCTP-14-43}

\title{On $\mathcal{N}=1$ partition functions without R-symmetry}

\author{Gino Knodel,}

\author{James T. Liu}

\author{and Leopoldo A. Pando Zayas}

\affiliation{Michigan Center for Theoretical Physics, Randall Laboratory
of Physics,\\
 The University of Michigan, Ann Arbor, MI 48109--1040, USA}

\emailAdd{gknodel@umich.edu} \emailAdd{jimliu@umich.edu} \emailAdd{lpandoz@umich.edu}

\abstract{We examine the dependence of four-dimensional Euclidean
$\mathcal{N}=1$ partition functions on coupling constants. In particular,
we focus on backgrounds without $R$-symmetry, which arise in the rigid
limit of old minimal supergravity. Backgrounds preserving a single
supercharge may be classified as having either chiral or non-chiral supercharges,
with the latter including $S^{4}$. We show that, in the absence of
additional symmetries, the partition function depends non-trivially
on all couplings in the non-chiral case, and (anti)-holomorphically
on couplings in the chiral case. In both cases, this allows
for ambiguities in the form of finite counterterms, which in principle render the partition function unphysical. However, we argue that on dimensional grounds, ambiguities are restricted to finite powers in relevant couplings, and can therefore be kept under control.
On the other hand, for backgrounds preserving two supercharges of opposite chiralities, the partition function is completely independent of all couplings.
In this case, the background admits an $R$-symmetry, and the partition
function is physical, in agreement with the results obtained in the
rigid limit of new minimal supergravity. Based on a systematic analysis
of supersymmetric invariants, we also demonstrate that $\mathcal N=1$ localization
is not possible for backgrounds without $R$-symmetry.} 

\maketitle
\tableofcontents{}

\section{Introduction and summary}

\label{sec:intro} Localization of supersymmetric field theories on
curved spaces has recently played a central role in elucidating some
long standing puzzles. Pestun made use of localization to compute
the expectation value of half supersymmetric Wilson loops in ${\cal N}=4$
supersymmetric Yang-Mills on $S^{4}$ and to prove that it is given
by a Gaussian matrix model \cite{Pestun:2007rz}, a conjecture made
more than a decade ago \cite{Drukker:2000rr,Erickson:2000af}. Kapustin,
Willett and Yaakov computed the partition function of supersymmetric
field theories on $S^{3}$ \cite{Kapustin:2009kz}, paving the way
to a better understanding of the number of degrees of freedom of such
theories, and clarifying various three-dimensional dualities (for a review, see e.g. \cite{Marino:2011nm}).

The program of computing supersymmetric observables on curved spaces
thus highlights the question of how to systematically construct such
field theories. Festuccia and Seiberg initiated a program to answer
this question in general, based on the principle of rigid supergravity
\cite{Festuccia:2011ws}. According to this principle, one considers
the field theory as a matter sector of a supergravity theory, and
then proceeds to decouple supergravity. The conditions for the background
to be supersymmetric are obtained by demanding that the gravitino
variation vanishes, and all couplings of the matter sector to the
background supergravity fields are dictated by the form of the supergravity
Lagrangian.

Rigid supergravity provides us with a powerful set of tools for answering
questions within a broad family of theories on curved spaces \cite{Jia:2011hw,Samtleben:2012gy,Klare:2012gn,Dumitrescu:2012ha,Liu:2012bi,deMedeiros:2012sb,Dumitrescu:2012at}.
One practically-minded question is whether we can perform localization
to calculate the partition function and other observables on various
curved backgrounds. Of particular interest are four-dimensional backgrounds
that do not possess an R-symmetry, such as the round and squashed
$S^{4}$, for which exact results for $\mathcal{N}=1$ theories have
so far been elusive. These backgrounds can be naturally studied in
the framework of old minimal supergravity, but they can not be found
as solutions to new minimal supergravity \cite{Sohnius:1981tp,Festuccia:2011ws,Closset:2013vra,Closset:2014uda,Dumitrescu:2012ha,Klare:2012gn}.

The standard localization procedure makes use of the fact that there
is at least one supersymmetric operator ${\cal O}$ such that the
partition function for a theory with Lagrangian ${\cal L}\supset t{\cal O}$
is independent of the corresponding coupling constant $t$, i.e. 
\begin{equation}
\frac{dZ(t)}{dt}=0.\label{eq:Zflat}
\end{equation}
If ${\cal O}$ has a positive semi-definite part, one can evaluate
the partition function at $t\gg1$, where it is given by a 1-loop
determinant around the classical configuration ${\cal O}_{cl}=0$.
To understand in which cases localization is in principle possible,
one needs to determine under which circumstances \eqref{eq:Zflat}
is satisfied. This condition is equivalent to the statement that there
is at least one ``flat'' direction in the space of coupling constants.
The first goal of this paper is therefore to better understand the
geometry of the space of couplings of ${\cal N}=1$ theories on four-dimensional
curved (Euclidean) backgrounds.

Our first step is to determine the supersymmetric invariants for a
given multiplet, which are the building blocks of supersymmetric Lagrangians.
In the standard approach, invariants are constructed using the tensor
calculus of supergravity \cite{Ferrara:1978jt,Ferrara:1978wj,Stelle:1978yr,Wess:1992cp},
and one finds the curved space generalizations of the flat-space $D$-term,
as well as a chiral $F$- and antichiral $\overline{F}$-term. However,
the supergravity approach assumes that the background manifold preserves
all four complex supercharges of Euclidean ${\cal N}=1$. As was shown
in \cite{Dumitrescu:2012ha,Dumitrescu:2012at,Liu:2012bi}, there is
a large set of interesting backgrounds with reduced supersymmetry,
which preserve fewer than four supercharges. In this case, there are
more than just the three standard SUSY-invariants. Furthermore, if the
background is $R$-symmetric, one may also combine superfields using
an antisymmetric product $S_{1}\wedge S_{2}$ to construct Lagrangians.

Since both of these subtleties are essentially invisible in the ``top-down''
approach of supergravity, we instead employ a ``bottom-up'' approach:
We take as our only input the curved space SUSY algebra, derived via
rigid supergravity \cite{Festuccia:2011ws,Jia:2011hw,Stelle:1978wj,Stelle:1978yr,Ferrara:1978jt,Wess:1992cp}.
Using the transformation rules, we can then construct the complete
set of SUSY invariants, as well as the multiplication rules for combining
supermultiplets. For a general Euclidean $\mathcal{N}=1$ multiplet
$S=(C,\psi_{L},\psi_{R},F,\overline{F},A_{\mu},\lambda_{L},\lambda_{R},D)$,
bosonic SUSY invariants take the form 
\begin{equation}
E=\alpha_{1}D+\alpha_{2}F+\alpha_{3}\overline{F}+\alpha_{4}C+\beta^{\mu}A_{\mu},
\end{equation}
with background-dependent coefficients $\alpha_{i},\beta^{\mu}$.
Demanding $E$ to be supersymmetric, we derive the differential conditions
on the coefficients and give examples of invariants.

A flat direction $t_{i}$ in the space of couplings is equivalent
to the statement that the corresponding invariant $E_{i}$ is $\delta$-exact.
One central result of this paper is that every invariant can be written
as a SUSY-exact term, plus extra terms that depend on the geometry
of the background. Schematically, we find 
\begin{equation}
E_{i}=\delta V_{i}+\xi_{i}^{\mu}A_{\mu}+\eta_{i}C,\label{eq:Ealm_exact}
\end{equation}
up to a total derivative, where $\xi_{i}^{\mu}$ and $\eta_{i}$ are
background-dependent. A flat direction exists only if $\xi_{i}^{\mu}=\eta_{i}=0$
for some $i$. We analyze Eqn.~\eqref{eq:Ealm_exact} for the backgrounds
of old minimal supergravity and extract properties of the space of
couplings. Our results can be summarized as follows: 
\begin{enumerate}
\item Backgrounds with non-chiral Killing spinors of the form $(\epsilon_{L},\epsilon_{R})$:
\\
 Such manifolds possess
an $S^{3}$-isometry (for example, the round and squashed $S^{4}$), but
do not admit an $R$-symmetry. We find that $\xi_{i}^{\mu},\eta_{i}\neq0$
for all invariants. This means that SUSY-closed terms are not exact,
and the partition function depends nontrivially on all coupling constants.
\item Backgrounds with chiral Killing spinors of the same chirality, i.e.
either $(\epsilon_{L},0)$ or $(0,\epsilon_{R})$: \\
Manifolds of this kind are characterized by $SU(2)_{R}$ or $SU(2)_{L}$
structure respectively, and possess a $U(1)_{R}$ $R$-symmetry. Focussing
on the former case, we find that all but one invariant are exact.
The exception is a generalized $\overline{F}$-term, so the partition
function only depends on the corresponding coupling%
\footnote{Considering instead backgrounds with $SU(2)_{L}$ structure amounts
to a flip of chiralities, so in this case there is a dependence on
${\lambda}_{F}$.%
} $\overline{\lambda}_{F}$. 
\item Backgrounds with chiral Killing spinors of opposite chirality, i.e.\ at
least one pair $(\epsilon_{L},0)$, $(0,\epsilon_{R})$: \\
 These are torus fibrations $T^{2}\times\Sigma$, where $\Sigma$
is a Riemann surface. Their structure group is reduced to the trivial
group, and there is a $U(1)_{R}$ $R$-symmetry. We find that all
invariants are exact, so the partition function is completely independent
of couplings. 
\end{enumerate}
In particular, we use our results to argue that localization of $\mathcal{N}=1$ theories on $S^{4}$
and the related cases in point 1 above is not possible: Since there
are simply no flat directions available, the Lagrangian cannot be deformed to perform localization. 
In the cases 2 and 3, localization proceeds in the
usual way. We give an explicit prescription for performing localization
on such backgrounds in section \ref{sub:SU2_loc}.

In the cases 1 and 2, the obvious question that arises is \textit{how}
the partition function depends on the couplings. The second goal of
this paper is therefore to analyze this dependence in detail, or in
other words, to determine which features of the space of couplings
are captured by the partition function. For certain superconformal
field theories (SCFTs), it was shown that the partition function computes
the Zamolodchikov metric on the space of exactly marginal couplings
\cite{Zamolodchikov:1986gt,Kutasov:1988xb,Gerchkovitz:2014gta,Jockers:2012dk,Gomis:2012wy,Doroud:2012xw,Benini:2012ui,Doroud:2013pka}.
Inspired by these results, we determine under which circumstances
one can extract similar physical quantities from $Z$. A generic complication
that arises is the fact that $Z$ itself does not always have an unambiguous
physical interpretation \cite{Assel:2014tba,Gerchkovitz:2014gta}: In general, finite
counterterms can shift the partition function according to 
\begin{equation}
\log Z\rightarrow\mathrm{log}Z+{\cal F}(\lambda_{i}),
\end{equation}
where ${\cal F}$ is a function of the couplings $\lambda_{i}$. If
such ambiguities are present, $Z$ is regularization scheme dependent,
and thus unphysical.

To determine the physical content of $Z$, it is therefore necessary
to classify the set of possible finite, supersymmetric counterterms.
Focussing on couplings to chiral/antichiral $F$- and $\overline{F}$-terms,
we perform a spurion analysis to construct such counterterms explicitly,
and determine whether or not they give rise to ambiguities in the
partition function. Let us again highlight some of our results: 
\begin{enumerate}
\item For backgrounds with non-chiral supercharges and $S^{3}$-isometry (i.e.\ backgrounds
without $R$-symmetry), there is an ambiguity of the form 
\begin{equation}
\mathrm{log}Z\sim\mathrm{log}Z+{F}(\lambda,\overline{\lambda})+G(\lambda)+H(\overline{\lambda}),
\end{equation}
where $F$, $G$ and $H$ are a priori unconstrained function of all
chiral/antichiral couplings $\lambda,\overline{\lambda}$. If we compute
$Z$ using different regularization schemes, we will find different
answers for its finite part, so the partition function itself is not
a sensible physical observable. However, if the theory contains relevant
couplings $m$, simple dimensional analysis reveals that the functions
$F$, $G$ and $H$ are in fact more constrained: They can only contain terms up to cubic order in $m$. We therefore
argue that all ambiguities can be removed by taking a suitable number of derivatives of $\mathrm{log}Z$
with respect to relevant couplings.
\item For backgrounds with $U(1)_{R}$ $R$-symmetry and $SU(2)_{R}$ structure,
the only nontrivial coupling is $\overline{\lambda}_F$. We find that the only ambiguity arises at quartic order in relevant couplings $\overline{m}$, and takes the form
\begin{equation}
\mathrm{log}Z\sim\mathrm{log}Z+b(\overline{m}r)^4,
\end{equation}
where $b$ is a background-dependent constant and $r$ is a characteristic length scale. 
\end{enumerate}
The rest of this paper is organized as follows. In section \ref{sec:rigidsusy},
we review the framework of rigid supersymmetry as applied to old minimal
supergravity, following the particular conventions and notation of
\cite{Liu:2012bi}. In section \ref{sec:trivial}, we discuss supersymmetric
theories on manifolds with non-chiral supercharges. We write down the general
$D$-type invariants, as well as the additional $F$-type invariants
for chiral superfields and analyze them in some detail. In particular,
we determine for which backgrounds they can be written as SUSY-exact
terms, so the partition function is independent of the corresponding
couplings. Using these results, we argue that partition functions
on manifolds with $S^{3}$-isometry depend nontrivially on all couplings,
which also implies that $\mathcal N=1$ localization is not possible. We proceed to discuss
the issue of ambiguities of the partition function by constructing
finite counterterms for chiral couplings. In section \ref{sec:SU2},
we analyze manifolds with a single chiral supercharge, or equivalently $SU(2)$ structure. We construct supersymmetric
invariants in an analogous way, and show that with one exception,
all SUSY-closed terms are also SUSY-exact, so $Z$ is again independent
of couplings. We then present the general philosophy of localization
using a simple toy model. We demonstrate that the dependence of $Z$
on antichiral $\overline{F}$-term couplings is ambiguous, and highlight
how the presence of a second supercharge of opposite chirality removes
the ambiguity completely, hence identifying torus fibrations as the only compact
backgrounds without ambiguities. Finally, we comment on explicit $R$-symmetry
breaking and the role of the auxiliary fields of supergravity. We
conclude with a discussion in section \ref{sec:discussion}.

\section{Rigid Supersymmetry\label{sec:rigidsusy}}

The general approach of rigid supersymmetry \cite{Festuccia:2011ws}
is to first start with a matter coupled supergravity theory and then
freeze out the gravitational sector, thus leaving a supersymmetric
field theory in a non-trivial background. Since we do not wish to
impose any gravitational dynamics on the background, it is necessary
to work in an off-shell formulation. In four dimensions, there are
two off-shell $\mathcal{N}=1$ supergravities --- one with the ``old
minimal'' set of auxiliary fields \cite{Stelle:1978ye,Ferrara:1978em}
and one with the ``new minimal'' set \cite{Sohnius:1981tp,Sohnius:1982fw}
--- both of which have been extended to the Euclidean case. Backgrounds
preserving an $R$-symmetry are naturally constructed in new minimal
supergravity, while those without $R$-symmetry only arise in old
minimal supergravity.

To avoid confusion about terminology, let us note that the theory we refer to as $\mathcal{N}=1$ possesses 4 real supercharges in Minkowski space. In Euclidean signature, one a priori has 4 complex supercharges, although certain backgrounds might break some of the supersymmetries. Our analysis does not apply to, for example, the SUSY theories on squashed 4-spheres considered in \cite{Nosaka:2013cpa}. Although these backgrounds admit either 2 or 4 supercharges, the SUSY algebra descends from a theory with 8 real supercharges in Minkowski space, i.e. $\mathcal{N}=2$.

\subsection{Supersymmetric backgrounds from old minimal supergravity}

The supergravity multiplet for off-shell supergravity with the ``old
minimal'' set of auxiliary fields is given by \cite{Stelle:1978ye,Ferrara:1978em,Wess:1992cp}
\begin{equation}
(g_{\mu\nu},\psi_{L\mu},\psi_{R\mu},b_{\mu},M,\overline{M}).
\end{equation}
In Euclidean signature, the chiral spinors $\psi_{L\mu}$ and $\psi_{R\mu}$
are independent, and transform under the left-/right-handed part of
$SO(4)=SU(2)_{L}\times SU(2)_{R}$. The auxiliary fields are a complex
vector $b_{\mu}$, and two independent complex scalars $M$, $\overline{M}$.

To find supersymmetric backgrounds, we assume a nontrivial background
metric $g_{\mu\nu}$, keeping the auxiliary fields arbitrary, but
set the gravitino and its variation equal to zero:

\begin{equation}
\delta\psi_{L\mu}=\delta\psi_{R\mu}=0.
\end{equation}
This condition gives rise to the following Killing spinor equations:
\begin{eqnarray}
\nabla_{\mu}\epsilon_{L} & = & \frac{1}{6}M\gamma_{\mu}\epsilon_{R}+\frac{i}{2}b_{\mu}\epsilon_{L}-\frac{i}{6}b^{\nu}\gamma_{\mu}\gamma_{\nu}\epsilon_{L},\nonumber \\
\nabla_{\mu}\epsilon_{R} & = & \frac{1}{6}\overline{M}\gamma_{\mu}\epsilon_{L}-\frac{i}{2}b_{\mu}\epsilon_{R}+\frac{i}{6}b^{\nu}\gamma_{\mu}\gamma_{\nu}\epsilon_{R}.\label{eq:KSE12}
\end{eqnarray}
A solution $\epsilon\equiv(\epsilon_{L},\epsilon_{R})$ corresponds
to a preserved supercharge. Generically, a background is specified
by an arbitrary configuration of the bosonic fields $(g_{\mu\nu},b_{\mu},M,\overline{M})$.
However, the condition that the background preserves supersymmetry
yields nontrivial constraints on the background fields. For each preserved
supercharge $\epsilon$, such a constraint is provided by the integrability
condition 
\begin{equation}
[\nabla_{\mu},\nabla_{\nu}]\epsilon=\frac{1}{4}R_{\mu\nu\lambda\sigma}\gamma^{\lambda\sigma}\epsilon,
\end{equation}
which relates the auxiliary fields $b_{\mu},M,\overline{M}$ to the
metric $g_{\mu\nu}$. A complete analysis of integrability conditions
in a case-by-case study was performed, for example, in \cite{Liu:2012bi,Dumitrescu:2012at}.
For our purposes, it is sufficient to note that demanding at least
one unbroken supersymmetry gives rise to the conditions 
\begin{eqnarray}
\gamma^{\mu}\nabla_{\mu}M\epsilon_{R} & = & \left(-\frac{1}{2}R+i\nabla_{\mu}b^{\mu}-\frac{2}{3}M\overline{M}-\frac{1}{3}b_{\mu}b^{\mu}\right)\epsilon_{L},\nonumber \\
\gamma^{\mu}\nabla_{\mu}\overline{M}\epsilon_{L} & = & \left(-\frac{1}{2}R-i\nabla_{\mu}b^{\mu}-\frac{2}{3}M\overline{M}-\frac{1}{3}b_{\mu}b^{\mu}\right)\epsilon_{R}.\label{eq:int12}
\end{eqnarray}
We can form a complete set of spinor bilinears that characterize the
background manifold: 
\begin{align}
f_{L} & =\epsilon_{L}^{\dagger}\epsilon_{L}, & f_{R} & =\epsilon_{R}^{\dagger}\epsilon_{R},\nonumber \\
Q_{\mu} & =\epsilon_{R}^{\dagger}\gamma_{\mu}\epsilon_{L}, & K_{\mu} & =\epsilon_{R}^{c}\gamma_{\mu}\epsilon_{L},\nonumber \\
J_{\mu\nu}^{L} & =i\epsilon_{L}^{\dagger}\gamma_{\mu\nu}\epsilon_{L}, & J_{\mu\nu}^{R} & =i\epsilon_{R}^{\dagger}\gamma_{\mu\nu}\epsilon_{R},\nonumber \\
\Omega_{\mu\nu}^{L} & =\epsilon_{L}^{c}\gamma_{\mu\nu}\epsilon_{L}, & \Omega_{\mu\nu}^{R} & =\epsilon_{R}^{c}\gamma_{\mu\nu}\epsilon_{R}.\label{eq:bilinears}
\end{align}
Throughout this paper, we follow the notation and conventions of \cite{Liu:2012bi}.

The existence of a nowhere vanishing Killing spinor $\epsilon$ imposes
additional structure on the supersymmetric backgrounds ${\cal M}$
considered here. There are two basic cases \cite{Liu:2012bi,Dumitrescu:2012at,Samtleben:2012gy,Lust:2010by}: 
\begin{itemize}
\item If the Killing spinor is of the form $(\epsilon_{L},\epsilon_{R})$,
with $f_{L}f_{R}\neq0$ (except at isolated points), the four
 vectors $Q_{\mu},Q_{\mu}^{\star},K_{\mu},K_{\mu}^{\star}$ provide a frame of linearly independent vectors (at least locally).
We refer to these manifolds as backgrounds
with non-chiral supercharges. They are discussed in section \ref{sec:trivial}. 
\item An interesting feature of Eqn.~(\ref{eq:KSE12}) is that a nowhere
vanishing solution $\epsilon$ still allows for either $\epsilon_{L}$
or $\epsilon_{R}$ to vanish identically, i.e. $f_{L}f_{R}=0$. Assuming
for concreteness that $\epsilon_{R}=0$, there are two linearly independent
spinors $\epsilon_{L}$ and $C\epsilon_{L}^{\star}$ characterizing
the background. Both spinors transform as singlets under $SU(2)_{R}$,
and the remaining structure group is $G=SU(2)_{R}$. Backgrounds with
$SU(2)$-structure are discussed in section \ref{sec:SU2}. 
\end{itemize}

\subsection{General multiplet and SUSY algebra}

\label{sec:gen_mult}

We now turn to the matter sector and its coupling to the supergravity
background. The general SUSY multiplet is given by \cite{Stelle:1978wj,Stelle:1978yr,Ferrara:1978jt}
\begin{equation}
S=(C,\psi_{L},\psi_{R},F,\overline{F},A_{\mu},\lambda_{L},\lambda_{R},D),
\end{equation}
and has $8+8$ components in Minkowski signature. In the Euclidean
case, the chiral spinors are taken to be independent, and all bosonic
fields are complex holomorphic variables. In particular, note that
$F$ and $\overline{F}$ are a priori independent, but will be related
to each other later by choosing an appropriate integration contour
in the path integral.

The curved space supersymmetry transformations of $S$ are found by
taking the rigid limit of the corresponding supergravity variations
\cite{Stelle:1978yr}: 
\begin{align}
\delta C & =-\epsilon_{L}^{c}\psi_{L}-\epsilon_{R}^{c}\psi_{R},\nn\\
\delta\psi_{L} & =\frac{1}{2}\gamma^{\mu}(A_{\mu}-\nabla_{\mu}C)\epsilon_{R}-\epsilon_{L}F,\nn\\
\delta\psi_{R} & =\frac{1}{2}\gamma^{\mu}(A_{\mu}+\nabla_{\mu}C)\epsilon_{L}+\epsilon_{R}\overline{F},\nn\\
\delta F & =\nabla^{\mu}\left(\epsilon_{R}^{c}\gamma_{\mu}\psi_{L}\right)-\overline{M}\epsilon_{L}^{c}\psi_{L}-\epsilon_{R}^{c}\lambda_{R},\nn\\
\delta\overline{F} & =\nabla^{\mu}\left(\epsilon_{L}^{c}\gamma_{\mu}\psi_{R}\right)-M\epsilon_{R}^{c}\psi_{R}-\epsilon_{L}^{c}\lambda_{L},\nn\\
\delta A_{\mu} & =\epsilon_{R}^{c}\gamma_{\mu}\lambda_{L}-\epsilon_{L}^{c}\gamma_{\mu}\lambda_{R}+\nabla_{\mu}\left(\epsilon_{L}^{c}\psi_{L}-\epsilon_{R}^{c}\psi_{R}\right),\nn\\
\delta\lambda_{L} & =\frac{1}{2}\gamma^{\mu\nu}\epsilon_{L}\nabla_{\mu}A_{\nu}-\frac{1}{2}\epsilon_{L}D,\nn\\
\delta\lambda_{R} & =\frac{1}{2}\gamma^{\mu\nu}\epsilon_{R}\nabla_{\mu}A_{\nu}+\frac{1}{2}\epsilon_{R}D,\nn\\
\delta D & =\nabla^{\mu}\left(\epsilon_{L}^{c}\gamma_{\mu}\lambda_{R}+\epsilon_{R}^{c}\gamma_{\mu}\lambda_{L}\right)+\frac{2i}{3}b_{\mu}\left(\epsilon_{L}^{c}\gamma^{\mu}\lambda_{R}-\epsilon_{R}^{c}\gamma^{\mu}\lambda_{L}\right)-\frac{2}{3}M\epsilon_{R}^{c}\lambda_{R}-\frac{2}{3}\overline{M}\epsilon_{L}^{c}\lambda_{L}.\label{eq:vec_susy19}
\end{align}
Irreducible representations can be embedded into $S$ by making certain
identifications \cite{Stelle:1978wj}. For example, a chiral multiplet
is given by 
\begin{equation}
\Phi=(\phi,\psi_{L},0,F,0,-\nabla_{\mu}\phi,0,0,0).\label{eq:chiral_embed}
\end{equation}
Similarly, an antichiral multiplet is embedded via 
\begin{equation}
\overline{\Phi}=(\overline{\phi},0,\psi_{R},0,\overline{F},\nabla_{\mu}\overline{\phi},0,0,0).\label{eq:antichiral_embed}
\end{equation}
The rules for multiplying two superfields $S_{1}$, $S_{2}$ are worked
out in the appendix. For SU(2) structure, there is an antisymmetric
product $S_{1}\wedge S_{2}$ in addition to the standard symmetric
product $S_{1}\times S_{2}$. This gives rise to some interesting
features when building supersymmetric Lagrangians (see section \ref{sub:SU2_loc}).

Throughout this paper, we take $\epsilon$ to be a commuting spinor
parameter. The closure relation of the algebra then takes the form
\begin{eqnarray}
\left\{ \delta_{1},\delta_{2}\right\}  & = & {\cal L}_{\xi},\label{eq:closure}
\end{eqnarray}
where ${\cal L}_{\xi}$ is the Lie derivative along the vector field
\begin{equation}
\xi^{\mu}=\epsilon_{1L}^{c}\gamma^{\mu}\epsilon_{2R}+\epsilon_{2L}^{c}\gamma^{\mu}\epsilon_{1R}.
\end{equation}
Since $\epsilon_{L}$ and $\epsilon_{R}$ transform independently
in Euclidean signature, the SUSY variation splits up into the action
of left- and right-handed components 
\begin{equation}
\delta=\delta_{L}+\delta_{R}=\epsilon_{L}^{c}Q_{L}+\epsilon_{R}^{c}Q_{R},
\end{equation}
corresponding to an anticommuting supercharge of the form $Q=(Q_{L},Q_{R})$.
Given this decomposition, we have 
\begin{equation}
\delta_{L}^{2}=\delta_{R}^{2}=0,\qquad\delta^{2}=\left\{ \delta_{L},\delta_{R}\right\} =-2{\cal L}_{K}.\label{eq:deltasq}
\end{equation}
While each $\delta_{L}$ and $\delta_{R}$ is nilpotent, the total
supercharge squares to a Lie derivative along the Killing vector $K^{\mu}$.
Since $K^{\mu}$ is in general complex, this provides an obstruction
to carrying out the usual localization procedure. Deforming the Lagrangian
by a SUSY-exact term $\sim\delta V$ generically breaks supersymmetry.
While this is an obvious complication for localization, it is not
sufficient to show that localization is not possible. One of the goals
in the remainder of this paper is to make the obstruction to localization
more precise, and provide a no-go theorem for localization on certain
manifolds with non-chiral supercharges.

One obvious way to avoid the above complication is to consider manifolds
with $SU(2)$ structure, where either $\epsilon_{L}$ or $\epsilon_{R}$
(and thus $K^{\mu}$) vanishes identically. In this case, $\delta$
is nilpotent and localization proceeds in the standard way. We analyze
this case in some detail in section~\ref{sec:SU2}.

\section{$\mathcal{N}=1$ theories on manifolds with non-chiral supercharges}

\label{sec:trivial}

We first consider the case of only  non-chiral supercharges, i.e. $f_{L}f_{R}\neq0$ (except at isolated points), because it allows us
to study manifolds that do not admit an $R$-symmetry. This includes,
in particular, the round and squashed $S^{4}$. In this case, the
space is spanned by four linearly independent vectors $Q_{\mu},Q_{\mu}^{\star},K_{\mu},K_{\mu}^{\star}$,
with $Q_{\mu}^{\star}Q^{\mu}=K_{\mu}^{\star}K^{\mu}=2f_{L}f_{R}$.
The two-forms in (\ref{eq:bilinears}) can be expressed in terms of
these vectors as 
\begin{align}
J^{L/R} & =-\frac{i}{2f_{R/L}}\left(K\wedge K^{\star}\pm Q\wedge Q^{\star}\right),\\
\Omega^{L} & =-\frac{1}{f_{R}}K\wedge Q,\\
\Omega^{R} & =\frac{1}{f_{L}}K\wedge Q^{\star}.
\end{align}
One can check that $\nabla_{(\mu}K_{\nu)}=0$, so that $K$ and $K^{\star}$
are Killing vectors. An interesting non-trivial feature in Euclidean
signature is that since $K$ and $K^{\star}$ are linearly independent,
their commutator may give rise to a third Killing vector 
\begin{equation}
L_{\mu}\equiv\left[K,K^{\star}\right]_{\mu}=\mu Q_{\mu}-\mu^{\star}Q_{\mu}^{\star},
\end{equation}
where 
\begin{equation}
\mu=\frac{1}{3}(f_{L}\overline{M}-f_{R}M^{\star})-\frac{2}{3}\mathrm{Im}(b^{\mu})Q_{\mu}^{\star}.
\end{equation}
Notice that $L$ is purely imaginary. The backgrounds then fall into
two different classes \cite{Liu:2012bi,Dumitrescu:2012at}: 
\begin{enumerate}
\item For $L\neq0$, the three Killing vectors $\mathrm{Re}K$, $\mathrm{Im}K$
and $L$ satisfy an $\mathfrak{su}(2)$-algebra, which allows us to
locally write the metric as a warped product $S^{3}\times\mathbb{R}$:
\begin{equation}
ds^{2}=d\xi^{2}+f(\xi)^{2}(\sigma_{1}^{2}+\sigma_{2}^{2}+\sigma_{3}^{2}).\label{eq:warped S3}
\end{equation}
Here $\sigma_{i}$ are the standard left-invariant one-forms on $S^{3}$.
Backgrounds of this form have been explicitly constructed \cite{Liu:2012bi,Dumitrescu:2012at}
and include the round $S^{4}$, $\mathbb{R}^{4}$, $\mathbb{H}^{4}$,
$S^{3}\times\mathbb{R}$ and $\mathbb{H}^{3}\times\mathbb{R}$, all
of which preserve four supercharges. Another interesting case is the
squashed $S^{4}$, which only preserves two supercharges. Notice that
at points where $f(\xi)=0$, either $\epsilon_{L}$ or $\epsilon_{R}$
vanishes. 
\item The case $L=0$ corresponds to a two-torus fibration over a Riemann
surface. This case splits up into two subclasses: 

\begin{enumerate}
\item The background has $M=\overline{M}=0$ and admits Killing spinors
of opposite chirality, namely $(\epsilon_{L},0)$ and $(0,\epsilon_{R})$.
This is equivalent to having independent left- and right-handed supercharges,
both of which are nilpotent. 
\item The Killing spinor has a chiral form $(\epsilon_{L},0)$ or $(0,\epsilon_{R})$.
This is the case of $SU(2)$ structure with a chiral supercharge (see
section \ref{sec:SU2}). 
\end{enumerate}
\end{enumerate}

\subsection{General invariants}

\label{sub:triv_gen_inv}In order to construct supersymmetric Lagrangians
on curved backgrounds, we will need the complete set of supersymmetric
invariants, which can be derived from the SUSY algebra. In the flat
space case with four supercharges, the bosonic invariants are the
usual $D$-terms and chiral $F$, $\overline{F}$-terms. Using the
tensor calculus for supergravity, these terms can be generalized to
curved space. The $D$-type invariant takes the form \cite{Wess:1992cp,Stelle:1978yr,Festuccia:2011ws,Jia:2011hw}
\begin{equation}
e^{-1}\int d^{2}\Theta(2\epsilon)(\overline{{\cal D}}{\cal \overline{D}}-8R)S=D+\frac{2}{3}(ib_{\mu}A^{\mu}-MF-\overline{M}F)-\left(\frac{1}{3}R-\frac{2}{9}M\overline{M}+\frac{2}{9}b_{\mu}b^{\mu}\right)C,\label{eq:gen_inv_sugra}
\end{equation}
where $2\epsilon=e\left(1-\Theta^{2}\overline{M}\right)$ is the chiral
density, $R$ is the curvature superfield, and $S$ is a general superfield.
In addition, there are generalized chiral $F$- and $\overline{F}$-terms
\begin{align}
e^{-1}\int d^{2}\Theta(2\epsilon)S & =F-\overline{M}\phi,\nonumber \\
e^{-1}\int d^{2}\overline{\Theta}(2\overline{\epsilon})S & =\overline{F}-M\overline{\phi}.
\end{align}
However, the superspace formalism generally assumes that the background
preserves the maximum number of supercharges. As we will demonstrate
in this section, relaxing the condition on the number or type of preserved
supercharges can give rise to additional invariants that are absent
in the top-down approach via supergravity. Hence we proceed with a
more systematic analysis of SUSY-invariants in curved space.

We consider a general superfield $S$, and make the following ansatz
for bosonic invariants: 
\begin{equation}
E=\alpha_{1}D+\alpha_{2}F+\alpha_{3}\overline{F}+\alpha_{4}C+\beta^{\mu}A_{\mu}.\label{eq:inv_general}
\end{equation}
We generally expect the coefficients $\alpha_{i}$, $\beta^{\mu}$
to be given in terms of the background fields $(g_{\mu\nu},b_{\mu},M,\overline{M})$.
However, as we will show, in some cases this restriction is too strong
(see section~\ref{sub:SU2_breakR}), so we treat them as a priori
arbitrary functions of $x$. On a compact manifold%
\footnote{For non-compact manifolds, one may impose suitable fall-off conditions
at infinity.%
}, $E$ is invariant if $\delta E$ is a total derivative. Assuming
no special field content (such as chiral/anti-chiral fields), this
gives rise to the following conditions: 
\begin{eqnarray}
(-\nabla^{\mu}\alpha_{1}+\frac{2i}{3}\alpha_{1}b^{\mu}-\beta^{\mu})\gamma_{\mu}\epsilon_{L}-(\frac{2}{3}\alpha_{1}M+\alpha_{2})\epsilon_{R} & = & 0,\nn\\
(-\nabla^{\mu}\alpha_{1}-\frac{2i}{3}\alpha_{1}b^{\mu}+\beta^{\mu})\gamma_{\mu}\epsilon_{R}-(\frac{2}{3}\alpha_{1}\overline{M}+\alpha_{3})\epsilon_{L} & = & 0,\nn\\
\nabla^{\mu}\alpha_{2}\gamma_{\mu}\epsilon_{R}+(\alpha_{2}\overline{M}+\nabla^{\mu}\beta_{\mu}+\alpha_{4})\epsilon_{L} & = & 0,\nn\\
\nabla^{\mu}\alpha_{3}\gamma_{\mu}\epsilon_{L}+(\alpha_{3}M-\nabla^{\mu}\beta_{\mu}+\alpha_{4})\epsilon_{R} & = & 0.\label{eq:inv_cond14}
\end{eqnarray}
Away from isolated points where one of the chiral Killing spinors
might vanish, we can write down a formal solution to this system of
equations: 
\begin{eqnarray}
\alpha_{2} & = & -(\nabla^{\mu}\alpha_{1}+\beta^{\mu}-\frac{2i}{3}\alpha_{1}b^{\mu})\frac{Q_{\mu}}{f_{R}}-\frac{2}{3}\alpha_{1}M,\nn\\
\alpha_{3} & = & (-\nabla^{\mu}\alpha_{1}+\beta^{\mu}-\frac{2i}{3}\alpha_{1}b^{\mu})\frac{Q_{\mu}^{\star}}{f_{L}}-\frac{2}{3}\alpha_{1}\overline{M},\nn\\
\alpha_{4} & = & -\frac{1}{2}\left(\alpha_{2}\overline{M}+\alpha_{3}M+\nabla^{\mu}\alpha_{2}\frac{Q_{\mu}^{\star}}{f_{L}}+\nabla^{\mu}\alpha_{3}\frac{Q_{\mu}}{f_{R}}\right),\nn\\
K^{\mu}\nabla_{\mu}\alpha_{1} & = & K^{\mu}\nabla_{\mu}\alpha_{2}=K^{\mu}\nabla_{\mu}\alpha_{3}=K^{\mu}(\beta_{\mu}-\frac{2i}{3}\alpha_{1}b_{\mu})=0,\nn\\
\nabla^{\mu}\beta_{\mu} & = & \frac{1}{2}\left(\nabla^{\mu}\alpha_{3}\frac{Q_{\mu}}{f_{R}}-\nabla^{\mu}\alpha_{2}\frac{Q_{\mu}^{\star}}{f_{L}}+\alpha_{3}M-\alpha_{2}\overline{M}\right).\label{eq:cond_scalar16}
\end{eqnarray}
For given functions $\alpha_{1}$ and $\beta^{\mu}$, the first three
equations in (\ref{eq:cond_scalar16}) determine $\alpha_{2}$, $\alpha_{3}$
and $\alpha_{4}$, respectively. The final two equations can then
be viewed as constraints on the form of $\alpha_{1}$ and $\beta^{\mu}$.

It is in general nontrivial to find solutions to the above system.
However, the analysis simplifies in the case of four supercharges.
Since, we can construct four linearly independent vectors $K_{i}^{\mu}$
(and similarly for $Q_{i}^{\mu}$), we conclude that $\alpha_{1}=\mathrm{constant}$.
Hence the only solution is%
\footnote{Note that this is an invariant even in the vicinity of isolated zeroes
of $\epsilon_{L/R}$. For a general invariant, one would need to
check this explicitly by plugging the solution to (\ref{eq:cond_scalar16})
back into (\ref{eq:inv_cond14}).%
} 
\begin{equation}
E\equiv D+\frac{2}{3}(ib_{\mu}A^{\mu}-MF-\overline{M}\overline{F}+M\overline{M}C),\label{eq:E4scharges}
\end{equation}
up to a constant rescaling. Using the integrability conditions (\ref{eq:int12}),
one can check that this is in fact a special case of the standard
$D$-type invariant \eqref{eq:gen_inv_sugra} of supergravity. For
backgrounds with less than maximal supersymmetry, there may be additional
solutions to (\ref{eq:cond_scalar16}), and hence more SUSY invariants.
We will not attempt to write down all invariants, but content ourselves
with giving some examples of additional invariants that arise in the
case of SU(2) structure in section \ref{sub:SU2_gen}.

We can nevertheless study the dependence of the partition function
on couplings to general invariants, without making use of explicit
solutions for $\alpha_{i},\beta_{\mu}$. An obvious question that
arises is whether or not $E$ can be written as a SUSY-exact term.
If this were true, the partition function would then be independent
of the coupling to such terms. To proceed, we again assume that $\epsilon_{L/R}\neq0$.
This allows us to rewrite the fermionic variations in (\ref{eq:vec_susy19})
as eight scalar equations by contracting with $\epsilon_{L/R}^{\dagger}$
and $\epsilon_{L/R}^{c}$: 
\begin{eqnarray}
F & = & -\delta\left(\frac{\epsilon_{L}^{\dagger}\psi_{L}}{f_{L}}\right)+\frac{1}{2f_{L}}Q_{\mu}^{\star}(A^{\mu}-\nabla^{\mu}C),\nn\\
\overline{F} & = & \delta\left(\frac{\epsilon_{R}^{\dagger}\psi_{R}}{f_{R}}\right)-\frac{1}{2f_{R}}Q_{\mu}\left(A^{\mu}+\nabla^{\mu}C\right),\nn\\
K^{\mu}A_{\mu} & = & \delta(\epsilon_{R}^{c}\psi_{R}-\epsilon_{L}^{c}\psi_{L}),\nn\\
K^{\mu}\nabla_{\mu}C & = & \delta(\epsilon_{R}^{c}\psi_{R}+\epsilon_{L}^{c}\psi_{L}),\nn\\
D & = & \delta\left(\frac{\epsilon_{R}^{\dagger}\lambda_{R}}{f_{R}}-\frac{\epsilon_{L}^{\dagger}\lambda_{L}}{f_{L}}\right)+(\frac{i}{2f_{R}}J_{\mu\nu}^{R}-\frac{i}{2f_{L}}J_{\mu\nu}^{L})\nabla^{\mu}A^{\nu},\nn\\
(\frac{i}{2f_{R}}J_{\mu\nu}^{R}+\frac{i}{2f_{L}}J_{\mu\nu}^{L})\nabla^{\mu}A^{\nu} & = & -\delta\left(\frac{\epsilon_{R}^{\dagger}\lambda_{R}}{f_{R}}+\frac{\epsilon_{L}^{\dagger}\lambda_{L}}{f_{L}}\right),\nn\\
\Omega_{\mu\nu}^{L}\nabla^{\mu}A^{\nu} & = & \delta(2\epsilon_{L}^{c}\lambda_{L}),\nn\\
\Omega_{\mu\nu}^{R}\nabla^{\mu}A^{\nu} & = & \delta(2\epsilon_{R}^{c}\lambda_{R}).\label{eq:exact18}
\end{eqnarray}
Using these relations along with (\ref{eq:cond_scalar16}) and the
integrability conditions (\ref{eq:int12}), we find that the general
invariant (\ref{eq:inv_general}) reduces to 
\begin{equation}
E=\delta V_{E}+\xi^{\mu}A_{\mu}+\eta C+\nabla(...),\label{eq:Easdelta}
\end{equation}
where 
\begin{equation}
V_{E}=\alpha_{1}\left(\frac{\epsilon_{R}^{\dagger}\lambda_{R}}{f_{R}}-\frac{\epsilon_{L}^{\dagger}\lambda_{L}}{f_{L}}\right)-\alpha_{2}\left(\frac{\epsilon_{L}^{\dagger}\psi_{L}}{f_{L}}\right)+\alpha_{3}\left(\frac{\epsilon_{R}^{\dagger}\psi_{R}}{f_{R}}\right)+\frac{2}{3f_{L}f_{R}}(\mathrm{Im}b\cdot K^{\star})\left(\epsilon_{L}^{c}\psi_{L}-\epsilon_{R}^{c}\psi_{R}\right),
\end{equation}
and 
\begin{align}
\xi_{\mu} & =\frac{1}{(2f_{L}f_{R})^{2}}\alpha_{1}(Q_{\nu}^{\star}Q_{\mu}-Q_{\nu}Q_{\mu}^{\star})L^{\nu},\nn\\
\eta & =\frac{1}{6f_{L}f_{R}}\alpha_{1}(ML\cdot Q^{\star}+\overline{M}L\cdot Q)+\frac{1}{2f_{L}f_{R}}L^{\mu}\nabla_{\mu}\alpha_{1}+\xi^{\mu}(\beta_{\mu}-\frac{2i}{3}\alpha_{1}b_{\mu}),
\end{align}
and $\nabla(...)$ denotes total derivatives. We see that in general,
$E$ cannot be written as a SUSY-exact term: There is an obstruction
in the form of additional terms that depend on the geometry. 
\begin{itemize}
\item Assuming $\alpha_{1}\neq0$, the extra terms vanish if and only if
$L=[K,K^{\star}]=0$, which is the case of torus fibrations. For $L=0$,
we then have two options: 

\begin{itemize}
\item If both $\epsilon_{L}$ and $\epsilon_{R}$ are nowhere vanishing,
$E=\delta V_{E}$ holds everywhere. We conclude that all SUSY invariants
are exact, and the partition function does not depend on the corresponding
couplings. This result is not surprising: As we noticed earlier, this
case corresponds to a pair of nilpotent supercharges $\delta_{L}^{2}=\delta_{R}^{2}=0$. 
\item If, for example, $\epsilon_{R}=0$ (which implies $G=SU(2)_{R}$),
the invariants can be written as a variation with respect to the left-handed
supercharge, $E=\delta_{L}V_{E}$ . We discuss this case in more detail
in section \ref{sub:SU2_gen}. Here we only note that the partition
function will again be independent of the couplings. 
\end{itemize}
\item For $L\neq0$, which includes the interesting case of $S^{4}$, equation
(\ref{eq:Easdelta}) demonstrates that there is no SUSY invariant
that is also exact, and hence we expect $Z$ to depend nontrivially
on all coupling constants. We analyze this dependence further in section
\ref{sub:triv_counter}, where we discuss the issue of finite counterterms. 
\end{itemize}
There is one invariant that needs to be discussed separately. Choosing
$\alpha_{1}=0$, Eqns. (\ref{eq:cond_scalar16}) imply that $\alpha_{2}=\alpha_{3}=\alpha_{4}=0$
and $\beta^{\mu}\sim K^{\mu}$. This corresponds to 
\begin{equation}
K^{\mu}A_{\mu}=\delta\left(\epsilon_{R}^{c}\psi_{R}-\epsilon_{L}^{c}\psi_{L}\right),\label{eq:KA_inv}
\end{equation}
which is SUSY-exact. This invariant generically only conserves a single
supercharge. We will further comment on the relevance of this term
in section \ref{sub:triv_loc}.

\subsection{Chiral invariants}

\label{sub:triv_chir_inv}In our analysis so far, we assumed that
there are no restrictions on the field content. Of course, any realistic
theory will have such restrictions. For example, a theory with chiral
and antichiral fields will admit generalized $F$-type and $\overline{F}$-type
invariants, in addition to the general $D$-type invariants (\ref{eq:inv_general}).

To find these additional chiral/antichiral invariants, we proceed
in a similar fashion as before. Chiral and antichiral multiplets are
embedded into the general multiplet as in (\ref{eq:chiral_embed})
and (\ref{eq:antichiral_embed}). The SUSY variations for a chiral
multiplet are 
\begin{align}
\delta\phi & =-\epsilon_{L}^{c}\psi_{L},\nn\\
\delta\psi_{L} & =-\gamma^{\mu}\epsilon_{R}\nabla_{\mu}\phi-\epsilon_{L}F,\nn\\
\delta F & =\nabla^{\mu}\left(\epsilon_{R}^{c}\gamma_{\mu}\psi_{L}\right)-\overline{M}\epsilon_{L}^{c}\psi_{L},\label{eq:chiral_susy13}
\end{align}
while for an antichiral multiplet, we have 
\begin{align}
\delta\overline{\phi} & =-\epsilon_{R}^{c}\psi_{R},\nn\\
\delta\psi_{R} & =\gamma^{\mu}\epsilon_{L}\nabla_{\mu}\overline{\phi}+\epsilon_{R}\overline{F},\nn\\
\delta\overline{F} & =\nabla^{\mu}\left(\epsilon_{L}^{c}\gamma_{\mu}\psi_{R}\right)-M\epsilon_{R}^{c}\psi_{R}.\label{eq:achiral_susy3}
\end{align}
The most general bosonic chiral/antichiral invariant may be written
as 
\begin{align}
I & =\beta_{1}F+\beta_{2}\phi,\nn\\
\overline{I} & =\overline{\beta}_{1}\overline{F}+\overline{\beta}_{2}\overline{\phi},\label{eq:IIbar_general}
\end{align}
with functions $\beta_{1}$, $\beta_{2}$, $\overline{\beta}_{1}$
and $\overline{\beta}_{2}$ to be determined. Demanding SUSY-invariance
of $I$ and $\overline{I}$ yields the conditions 
\begin{align}
\nabla^{\mu}\beta_{1}\gamma_{\mu}\epsilon_{R}+(\beta_{2}+\beta_{1}\overline{M})\epsilon_{L} & =0,\nn\\
\nabla^{\mu}\overline{\beta}_{1}\gamma_{\mu}\epsilon_{L}+(\overline{\beta}_{2}+\overline{\beta}_{1}M)\epsilon_{R} & =0,\label{eq:cond_chiral12}
\end{align}
or equivalently 
\begin{align}
\beta_{2} & =-\beta_{1}\overline{M}-\nabla^{\mu}\beta_{1}\frac{Q_{\mu}^{\star}}{f_{L}},\nn\\
\overline{\beta}_{2} & =-\overline{\beta}_{1}M-\nabla^{\mu}\overline{\beta}_{1}\frac{Q_{\mu}}{f_{R}},\nn\\
K^{\mu}\nabla_{\mu}\beta_{1} & =K^{\mu}\nabla_{\mu}\overline{\beta}_{1}=0.
\end{align}
Again, for a background that preserves four supercharges, the only
solution is to take $\beta_{1}$, $\overline{\beta}_{1}$ to be constants,
so the invariants are 
\begin{align}
I & =F-\overline{M}\phi,\nn\\
\overline{I} & =\overline{F}-M\overline{\phi}.\label{eq:IIbar4sc}
\end{align}
These are the curved space generalization of the standard $F$, $\overline{F}$-terms.
The coupling to the background fields can be thought of as originating
from the nontrivial chiral density $2\epsilon$ in the superspace
formalism: 
\begin{equation}
S\big|_{I}=e^{-1}\int d^{2}\Theta(2\epsilon)S,\qquad S\big|_{\overline{I}}=e^{-1}\int d^{2}\overline{\Theta}(2\overline{\epsilon})\overline{S}.
\end{equation}
After setting the gravitino to zero, we find $2\epsilon=e\left(1-\Theta^{2}\overline{M}\right)$,
which shifts the $F$-terms as in (\ref{eq:IIbar4sc}).

For backgrounds that preserve fewer than four supercharges, there
may be more solutions to (\ref{eq:cond_chiral12}). We can ask if
a general invariant $I,\overline{I}$, with $\beta_{1},\overline{\beta}_{1}$
unspecified, is SUSY-exact. We find that 
\begin{align}
I & =\delta\left(-\beta_{1}\frac{\epsilon_{L}^{\dagger}\psi_{L}}{f_{L}}\right)-\frac{1}{2f_{L}^{2}f_{R}}\beta_{1}(L\cdot Q^{\star})\phi+\nabla(...),\nn\\
\overline{I} & =\delta\left(\overline{\beta}_{1}\frac{\epsilon_{R}^{\dagger}\psi_{R}}{f_{R}}\right)-\frac{1}{2f_{L}f_{R}^{2}}\overline{\beta}_{1}(L\cdot Q)\overline{\phi}+\nabla(...).\label{eq:IIbar_exact}
\end{align}
As before, the obstruction to exactness is related to $K^{\mu}$ not
commuting with its complex conjugate. 
\begin{itemize}
\item For backgrounds with $S^{3}$-isometry, where $L\neq0$, neither $I$
nor $\overline{I}$ are exact, and the partition function depends
nontrivially on all chiral/antichiral couplings. 
\item For torus fibrations, where $L=0$, $I$ and $\overline{I}$ are in
general exact, and the partition function is independent of chiral/antichiral
couplings. Notice however that if one of the chiral spinors $\epsilon_{L}$
or $\epsilon_{R}$ vanishes identically, then either the $I$ or $\overline{I}$
equation in (\ref{eq:IIbar_exact}) is no longer valid. We discuss
this case separately in section \ref{sub:SU2_chir}. 
\end{itemize}

\subsection{Lagrangians and localization}

\label{sub:triv_loc}With the knowledge of the SUSY invariants, one
can construct Lagrangians for an arbitrary field content. As an instructive
example, we will discuss the case of a chiral and antichiral multiplet
$(\Phi,\overline{\Phi})$. Guided by the ``no-miracles'' principle,
we should write down the most general terms consistent with the symmetries
of the theory. We have seen that the invariants are the D-type terms
(\ref{eq:inv_general}), and the chiral/antichiral F-type invariants
(\ref{eq:IIbar_general}). Hence the most general Lagrangian is

\begin{equation}
e^{-1}{\cal L}=-\frac{1}{2}\sum_{E}K(\Phi,\overline{\Phi})\bigg|_{E}-\sum_{I}W(\Phi)\bigg|_{I}-\sum_{\overline{I}}\overline{W}(\overline{\Phi})\bigg|_{\overline{I}}.\label{eq:Ltrivial}
\end{equation}
Here $K$ is a K{\"a}hler potential, which can be written as a power
series involving the $\times$-multiplication (see appendix \ref{sec:tens_calc}),
and $W$ is the holomorphic superpotential. The sums are taken over
all possible invariants for a given background. For a maximally supersymmetric
space, there are only three invariants, namely the $E$ invariant
of (\ref{eq:gen_inv_sugra}), and the $I$ and $\overline{I}$ invariants
of (\ref{eq:IIbar4sc}), so the analysis simplifies somewhat. In this
case, evaluating (\ref{eq:Ltrivial}) yields 
\begin{eqnarray}
e^{-1}{\cal L} & = & K\left(\frac{1}{6}R+\frac{1}{9}b^{\mu}b_{\mu}-\frac{1}{9}M\overline{M}\right)+K^{(1,1)}\left(\partial^{\mu}\bar{\phi}\partial_{\mu}\phi-F\overline{F}\right)+\frac{i}{3}b^{\mu}\left(K^{(1,0)}\partial_{\mu}\phi-K^{(0,1)}\partial_{\mu}\bar{\phi}\right)\nonumber \\
 &  & +F\left(\frac{1}{3}MK^{(1,0)}-W^{(1)}\right)+\bar{F}\left(\frac{1}{3}\overline{M}K^{(0,1)}-\overline{W}^{(1)}\right)+W\overline{M}+\overline{W}M\nonumber \\
 &  & +K^{(1,1)}\psi_{R}^{c}\gamma^{\mu}\widetilde{\nabla}_{\mu}\psi_{L}+\frac{1}{2}\left(W^{(2)}+K^{(2,1)}\overline{F}-\frac{1}{3}MK^{(2,0)}\right)\psi_{L}^{c}\psi_{L}\nonumber \\
 &  & +\frac{1}{2}\left(\overline{W}^{(2)}+K^{(1,2)}F-\frac{1}{3}\overline{M}K^{(0,2)}\right)\psi_{R}^{c}\psi_{R}-\frac{1}{4}K^{(2,2)}\psi_{L}^{c}\psi_{L}\psi_{R}^{c}\psi_{R},\label{eq:Lfull}
\end{eqnarray}
where $K^{(n,m)}\equiv{\partial^{n+m}}K/{\partial\phi^{n}\partial\overline{\phi}^{m}}$
and $W^{(n)}\equiv{\partial^{n}}W/{\partial\phi^{n}}$. We have also
defined 
\begin{equation}
\widetilde{\nabla}_{\mu}\psi_{L}=\left(\nabla_{\mu}+\frac{i}{6}K^{(1,1)}b_{\mu}+K^{(1,1)}K^{(2,1)}\partial_{\mu}\phi\right)\psi_{L}.
\end{equation}
The Lagrangian (\ref{eq:Lfull}) is of course the same result one
obtains by taking the rigid limit of the supergravity Lagrangian,
analytically continued to Euclidean signature \cite{Festuccia:2011ws,Jia:2011hw,Wess:1992cp}.

An interesting question is whether the partition function for (\ref{eq:Ltrivial})
can be computed via localization. Let us therefore review the general
philosophy of localization \cite{Pestun:2007rz,Kapustin:2009kz}.
Given a supersymmetric Lagrangian, one considers a deformation of the original theory,
\begin{equation}
{\cal L}\rightarrow{\cal L}+t{\cal L}_{t}
\end{equation}
such that the partition function remains invariant, i.e.
\begin{equation}
\frac{dZ(t)}{dt}=0.\label{eq:Zindep}
\end{equation}
We can then evaluate $Z$ for any given value of $t$, and are guaranteed
to get the same result. In particular, we can go to a corner in the
space of couplings where $t$ is much bigger than all other coupling constants in the theory (i.e.\ formally take
$t\rightarrow\infty$), and compute $Z$ there. If ${\cal L}_{t}$
has a positive semi-definite bosonic part, the theory localizes around
the classical locus ${\cal L}_{t}\big|_{\mathrm{bos.}}=0$ and the
partition function is one-loop exact.

The necessary and sufficient condition for (\ref{eq:Zindep}) to hold
is that the corresponding deformation ${\cal{L}}_{t}$ of the Lagrangian is exact with respect to one of the supercharges: 
\begin{equation}
{\cal L}_{t}=\delta V.
\end{equation}
 Localization then simply utilizes the fact that the deformed theory posseses a ``flat'' direction
in the space of coupling constants.

To summarize, there are two basic conditions that need to be satisfied
for localization: 
\begin{enumerate}
\item There exists a deformation ${\cal L}_{t}$ that is both SUSY-closed
and exact with respect to (at least) one supercharge. 
\item The bosonic part of ${\cal L}_{t}$ is positive semi-definite. 
\end{enumerate}
Using our results from the previous section, we can easily check these
conditions for a broad class of manifolds. 
\begin{itemize}
\item For the case of non-chiral supercharges with $L=[K,K^{\star}]=0$, all SUSY invariants
are exact. In principle, there is no obstruction to performing localization.
The nontrivial task is to find a positive semi-definite localization
term. We will do so for the closely related case of manifolds with
SU(2) structure in section \ref{sub:SU2_loc}. 
\item For $L\neq0$, there are no terms that are both SUSY-closed and exact. We conclude that
for manifolds with $S^{3}$-isometry, in particular the squashed and
round $S^{4}$, the partition function does not localize. 
\end{itemize}
Finally, there is one invariant that needs to be discussed separately,
namely (\ref{eq:KA_inv}), which preserves only one supercharge. Evaluated
on a K{\"a}hler potential $K(\Phi,\overline{\Phi})$, it reads 
\begin{equation}
K^{\mu}A_{\mu}=K^{\mu}\left(K^{(1,0)}\nabla_{\mu}\overline{\phi}-K^{(0,1)}\nabla_{\mu}\phi+K^{(1,1)}\psi_{L}^{c}\gamma_{\mu}\psi_{R}\right).\label{eq:KAonK}
\end{equation}
This can be regarded as a coupling of the global $U(1)$-current to
the background. It is obvious that its bosonic part cannot be made
positive semi-definite, so (\ref{eq:KAonK}) cannot be used for localization.

We should note that our result strictly speaking only holds for a
chiral/antichiral field content. Considering other irreducible representations,
such as gauge or linear multiplets, might lead to additional invariants,
analogous to the chiral/antichiral $I$ and $\overline{I}$ terms.
However, our analysis of the general $D$-type terms was independent
of the field content, so it is still true that for $L\neq0$, there
are no $D$-type invariants that are exact. Since the kinetic terms
for fields are generally only found among these $D$-terms, we conjecture
that the possible additional invariants cannot be utilized for localization.

\subsection{Counterterms and the physical part of $Z$}

\label{sub:triv_counter}For the case $L\neq0$, we have established
that the partition function is a nontrivial function of all couplings
(with one exception, see above). The next question to ask is whether this
dependence is non-ambiguous.

It is instructive to review the logic of extracting physical data
from partition functions. On compact manifolds, infrared divergences
in the partition function are absent, due to the finite volume of the background. However, there might still be
ultraviolet divergences that need to be regularized. Very schematically,
the partition function may take the form 
\begin{equation}
\mathrm{log}Z(\lambda_{i})=\sum_{j}a_{j}(\lambda_{i})\Lambda^{j}+A(\lambda_{i})\mathrm{log}\Lambda+F\left(\lambda_{i}\right).\label{eq:logF_schem}
\end{equation}
The first term captures power law divergences, with $\Lambda$ being
the UV cutoff. The log-divergent term is the analog of the A-type
anomaly in CFTs. The last part is the finite contribution $F$ to
the free energy. From our analysis above, we expect all terms to be
nontrivial functions of the couplings.

A regularization scheme corresponds to choosing a certain set of counterterms,
which can be used to tune some of the terms in (\ref{eq:logF_schem})
to zero. Only the parts of the partition function that are unaffected
by counterterms are physical observables%
\footnote{Note that in some cases one might have to take a certain number of
derivatives of $\mathrm{log}Z$ with respect to the couplings $\lambda_{i}$
to extract the unambiguous physical data. One example is the Zamolodchikov
metric on the space of exactly marginal couplings of a CFT \cite{Zamolodchikov:1986gt,Kutasov:1988xb,Gerchkovitz:2014gta,Jockers:2012dk,Gomis:2012wy,Doroud:2012xw,Benini:2012ui,Doroud:2013pka},
$g_{i\overline{j}}\sim\partial_{i}\partial_{\overline{j}}\mathrm{log}Z(\lambda_{k},\overline{\lambda}_{l})$.%
}. Let us now determine the physical content of ${\cal N}=1$ theories
on backgrounds with $L^{\mu}\neq0$. Instead of considering all possible
couplings, we focus on couplings to chiral invariants $I$,$\overline{I}$. The
interactions take the form 
\begin{equation}
e^{-1}{\cal L}_{\mathrm{int}}=W\big|_{I}+\overline{W}\big|_{\overline{I}}.\label{eq:chiral_invs}
\end{equation}
For a renormalizable theory, the superpotential $W$ contains relevant
couplings $m_{i}$ and marginal couplings $\lambda_{i}$. We can classify
the possible counterterms by performing a spurion analysis. For clarity
of presentation, we will focus on the case of only a single pair of
relevant couplings $(m,\overline{m})$ and marginal couplings $(\lambda,\overline{\lambda})$ each.
The generalization to an arbitrary number of couplings
should be straightforward. Treating the couplings as the lowest components
of spurious  chiral/antichiral superfields $({\Sigma_m},{\overline{\Sigma}_m},\Sigma_{\lambda},\overline{\Sigma}_{\lambda})$,
we see that renormalizable interactions arise from 
\begin{equation}
e^{-1}{\cal L}_{\mathrm{int}}=\big[{\Sigma_m}\Phi^{2}+\Sigma_{\lambda}\Phi^{3}\big]\big|_{I}+\big[\overline{\Sigma}_m\overline{\Phi}^{2}+\overline{\Sigma}_{\lambda}\overline{\Phi}^{3}\big]\big|_{\overline{I}}
\end{equation}
upon taking expectation values.

The possible finite counterterms are local interactions of spurions, consistent with the
symmetries of the underlying theory. Let us start by choosing a supersymmetric
background with the smallest possible set of symmetries. Those are
manifolds with only one conserved supercharge, so
the desired counterterms are all local, diffeomorphism invariant
terms that preserve one supersymmetry. We have already derived the
complete set of such terms, so we can conclude that the counterterms
are given by the $E$-, $I$-, and $\overline{I}$-terms of sections~\ref{sub:triv_gen_inv}
and \ref{sub:triv_chir_inv}. Hence the possible finite counterterms arise
from interactions of the form 
\begin{equation}
F(\Sigma_{\lambda},\overline{\Sigma}_{\lambda},{\Sigma_m},{\overline{\Sigma}_m})\big|_{E}+G(\Sigma_{\lambda},{\Sigma_m})\big|_{I}+H(\overline{\Sigma}_{\lambda},\overline{{\Sigma}}_m)\big|_{\overline{I}} .
\end{equation}
Taking the appropriate expectation values, we find the following counterterm
Lagrangian: 
\begin{equation}
e^{-1}{\cal L}_{ct}=\alpha_{4}F(\lambda,\overline{\lambda},m,\overline{m})+\beta_{2}G(\lambda,m)+\overline{\beta}_{2}H(\overline{\lambda},\overline{m}).\label{eq:ct_gen}
\end{equation}
Here $\alpha_{4}$, $\beta_{2}$ and $\overline{\beta}_{2}$ are solutions
to the system (\ref{eq:cond_scalar16}). Instead of attempting to work with the most general
solution, let us simply note that the standard choice
\begin{align}
\alpha_{4} & =-\frac{1}{3}R+\frac{2}{9}M\overline{M}-\frac{2}{9}b_{\mu}b^{\mu},\\
\beta_{2} & =-\overline{M},\\
\overline{\beta}_{2} & =-M,
\end{align}
is a solution for any number of preserved supercharges and work with
the invariants corresponding to this choice.

Using dimensional analysis, we can further constrain the form of the
counterterms. Assuming that $\Phi$ is canonically normalized, we
have $[m]=1$, so the function $F$ in (\ref{eq:ct_gen}) needs to be a quadratic function of relevant couplings, while $G$ and $H$ are cubic.
Carrying out the volume integral to compute the action will produce
a curvature scale $\int\sqrt{g}\alpha_{4}\sim r^{2}$, and similarly
for $\beta_{2}$, $\overline{\beta}_{2}$. Thus the partition function itself exhibits a regularization scheme
dependent ambiguity of the form 
\begin{equation}
\log Z\sim\log Z+f\left(mr,\overline{m}r,\lambda,\overline{\lambda}\right)+(mr)^{3}g(\lambda)+(\overline{m}r)^{3}h(\overline{\lambda}),\label{eq:ambiguity_general}
\end{equation}
where $f$ contains only terms that are quadratic in relevant couplings. To be completely general, we should also consider counterterms that
involve curvature multiplets \cite{Assel:2014tba,Gerchkovitz:2014gta}.
For example, there are D-type counterterms of the form

\begin{equation}
{\cal R}^{i}\overline{{\cal R}}^{j}{\Sigma}_m^{k}\overline{{\Sigma}}_m^{l}F(\Sigma_{\lambda},\overline{\Sigma}_{\lambda})\big|_{E},\label{eq:ct_examples}
\end{equation}
where ${\cal {R}}$ is the chiral curvature superfield, with expectation
value
\begin{equation}
-6\left<{\cal R}\right>=M+\Theta^{2}\left(\frac{1}{2}R+\frac{2}{3}M\overline{M}+\frac{1}{3}b^{\mu}b_{\mu}-i\nabla^{\mu}b_{\mu}\right).\label{eq:Rcurv}
\end{equation}
Since its lowest component has mass dimension 1, we see that $i+j+k+l=2$ in (\ref{eq:ct_examples}). In addition, we should also consider the more general chiral/antichiral counterterms

\begin{equation}
{\cal R}^{i}{\Sigma}_m^{3-i}G(\Sigma_{\lambda})\big|_{I},\quad{\cal \overline{R}}^{i}{\overline{\Sigma}}_m^{3-i}H(\overline{\Sigma}_{\lambda})\big|_{\overline{I}}.
\end{equation}
If we include all such mixed matter-gravity counterterms, the ambiguity becomes 
\begin{equation}
\log Z\sim\log Z+F_{2}\left(mr,\overline{m}r,\lambda,\overline{\lambda}\right)+G_{3}(mr,\lambda)+H_{3}(\overline{m}r,\overline{\lambda}),\label{eq:ambiguity_withR}
\end{equation}
where $F_{2}$,$G_{3}$,$H_{3}$ are now general quadratic (cubic) polynomials in
the relevant couplings, but arbitrary functions of marginal couplings:
\begin{align}
F_{2}\left(mr,\overline{m}r,\lambda,\overline{\lambda}\right) & =\sum_{i+j\leq2}a_{i,j}(mr)^{i}(\overline{m}r)^{j}f_{i,j}(\lambda,\overline{\lambda}),\nonumber \\
G_{3}(mr,\lambda) & =\sum_{i\leq3}b_{i}(mr)^{i}g_{i}(\lambda),\nonumber \\
H_{3}(\overline{m}r,\overline{\lambda}) & =\sum_{i\leq3}c_{i}(\overline{m}r)^{i}h_{i}(\overline{\lambda}).\label{eq:counter_polyn}
\end{align}
The coefficients $a,b,c$ are dimensionless, background-dependent
constants that arise from integrating curvature invariants. 

We conclude
that in general, finite counterterms may shift the free energy by
regularization scheme dependent terms according to (\ref{eq:ambiguity_withR}). 
If we expand $\mathrm{log}Z(m,\lambda)$ in powers of relevant couplings, all terms up to cubic order are subject to ambiguities, and thus unphysical. However, higher powers of $m$ are free from ambiguities, so we may extract the physical part of the partition function by 
taking suitable derivatives with respect to coupling
constants. Inspecting \eqref{eq:counter_polyn},
we see that, for example, 
\begin{equation}
\frac{\partial^{4}}{\partial(mr)^{4}}\mathrm{log}Z,\quad\frac{\partial^{3}}{\partial(mr)^{3}}\frac{\partial^{2}\mathrm{log}Z}{\partial\lambda\partial\overline{\lambda}},\label{eq:remove}
\end{equation}
are unambiguous physical observables. The minimum number of derivatives one has to take is model-dependent, since additional global symmetries may forbid certain counterterms. Note that the
second expression in (\ref{eq:remove}) is reminiscent of the Zamolodchikov metric for CFTs
\cite{Zamolodchikov:1986gt,Kutasov:1988xb,Gerchkovitz:2014gta,Jockers:2012dk,Gomis:2012wy,Doroud:2012xw,Benini:2012ui,Doroud:2013pka}.

Another way to avoid counterterm ambiguities of the partition
function is to further constrain the background manifolds, such that
the coefficients multiplying the counterterms in \eqref{eq:ct_gen}
vanish identically. The rule of thumb is that more symmetries imply
fewer counterterms, which allows for more physical observables to
exist. For manifolds with the maximum number of four preserved supercharges,
the integrability conditions (\ref{eq:int12}) imply 
\begin{equation}
R=-\frac{4}{3}M\overline{M}-\frac{2}{3}b^{\mu}b_{\mu},
\end{equation}
and hence
\begin{equation}
\alpha_{4}=\frac{2}{3}M\overline{M},\quad\beta_{2}=-\overline{M},\quad\overline{\beta}_{2}=-M.
\end{equation}
 Following \cite{Dumitrescu:2012at}, there are two types of backgrounds 
\begin{itemize}
\item For $M,\overline{M}\neq0$, the space is locally isometric to the
round $S^{4}$ or $\mathbb{H}^{4}$. In this case, the ambiguity (\ref{eq:ambiguity_withR})
remains. Since the round sphere is a limiting case of the squashed
sphere $\tilde{S}^{4}$, $\alpha_{4}$ cannot vanish identically for
$\tilde{S}^{4}$, so the ambiguity is present in this case as well.
\item For $M=\overline{M}=0$, the background is locally isometric to ${\cal M}_{3}\times\mathbb{R}$,
where ${\cal M}_{3}$ has constant curvature. In this case, the candidate
counterterms vanish identically, and there is no obvious obstruction
for the finite part of $F=\log Z$ to be a physical observable. To
prove that $F$ is indeed physical, one would need to perform a more complete analysis involving also purely gravitational counterterms constructed out of the curvature multiplets of supergravity,
along the lines of \cite{Assel:2014tba}. 
\end{itemize}
It is interesting to compare our result \eqref{eq:ambiguity_withR}
to the case of SCFTs on $S^{4}$ \cite{Gerchkovitz:2014gta}. In the
latter case, there is no mass scale $m$. However, counterterms that couple marginal operators to the background (e.g. the case $k=l=0$ in (\ref{eq:ct_examples})) are still present, so there is an ambiguity of
the form 
\begin{equation}
\mathrm{log}Z\sim\mathrm{log}Z+f(\lambda,\overline{\lambda}),
\end{equation}
where $f$ is an arbitrary function of the marginal couplings, and the finite part of the partition function is completely unphysical.


\section{$\mathcal{N}=1$ theories on manifolds with chiral supercharges}

\label{sec:SU2}


We now turn to the case of manifolds with supercharges of definite chirality, which possess SU(2)-structure and
an $R$-symmetry. In general, the Killing spinor equations (\ref{eq:KSE12})
mix left- and right-handed spinors, so there can be no $R$-symmetry.
However, this mixing is not present whenever either $\epsilon_{L}$
or $\epsilon_{R}$ vanish identically. Without loss of generality,
we will assume that there is a supercharge of the form $(\epsilon_{L},0)$.
Setting $\epsilon_{R}=0$ in (\ref{eq:KSE12}) then yields the Killing
spinor equation 
\begin{equation}
\nabla_{\mu}\epsilon_{L}=\frac{i}{2}b_{\mu}\epsilon_{L}-\frac{i}{6}b^{\nu}\gamma_{\mu}\gamma_{\nu}\epsilon_{L},\label{eq:KSE_SU2}
\end{equation}
along with the requirement $\overline{M}=0$. Note that $M$ has completely
dropped out of this expression, so a priori it is an arbitrary function.

The class of backgrounds described above possesses a $U(1)_{R}$ $R$-symmetry,
under which $\epsilon_{L}$ carries charge 1. Theories with R-symmetry
can be naturally coupled to new minimal supergravity \cite{Sohnius:1981tp,Festuccia:2011ws,Closset:2013vra,Closset:2014uda,Dumitrescu:2012ha,Klare:2012gn}.
In this framework, the six auxiliary degrees of freedom are captured
by a conserved vector $V_{\mu}$ and a $U(1)_{R}$-gauge field $A_{\mu}$.
The conditions for a background to preserve supersymmetry are \cite{Festuccia:2011ws}
\begin{eqnarray}
D_{\mu}\epsilon_{L} & = & -\frac{3i}{2}V_{\mu}\epsilon_{L}+\frac{i}{2}V^{\nu}\gamma_{\mu}\gamma_{\nu}\epsilon_{L},\nn\\
D_{\mu}\epsilon_{R} & = & \frac{3i}{2}V_{\mu}\epsilon_{R}-\frac{i}{2}V^{\nu}\gamma_{\mu}\gamma_{\nu}\epsilon_{R},
\end{eqnarray}
where $D_{\mu}=\nabla_{\mu}-ir(A_{\mu}+\frac{3}{2}V_{\mu})$ is an
$R$-covariant derivative. The left- and right-handed supercharges
carry $R$-charges $1$ and $-1$. If we restrict to a subclass of
backgrounds with 
\begin{equation}
A_{\mu}=-\frac{3}{2}V_{\mu},\qquad V_{\mu}\equiv-\frac{1}{3}b_{\mu},
\end{equation}
we recover (\ref{eq:KSE_SU2}) and its right-handed counterpart. Hence
backgrounds with SU(2) structure in old-minimal supergravity are a
subclass of the backgrounds of new minimal supergravity.

Let us briefly summarize some known features of the backgrounds ${\cal M}$
considered here. From (\ref{eq:KSE_SU2}), we can derive the integrability
conditions 
\begin{align}
R & =2i\nabla\cdot b-\frac{2}{3}b_{\mu}b^{\mu},\nn\\
\partial_{[\mu}b_{\nu]} & =\frac{1}{2}\epsilon_{\mu\nu\lambda\sigma}\partial^{\lambda}b^{\sigma}.
\end{align}
As before, we can construct bilinears from the Killing spinor: 
\begin{equation}
f_{L}=\epsilon_{L}^{\dagger}\epsilon_{L},\qquad J_{\mu\nu}=\frac{i}{f_{L}}\epsilon_{L}^{\dagger}\gamma_{\mu\nu}\epsilon_{L},\qquad\Omega_{\mu\nu}=\epsilon_{L}^{c}\gamma_{\mu\nu}\epsilon_{L}.
\end{equation}
Note that there are no invariant vectors in the SU(2) structure case.
Using Fierz identities, we have $J_{\mu\nu}J^{\nu\rho}=-\delta_{\mu}^{\rho}$,
so $J$ defines an almost complex structure. It can be shown that
the corresponding Nijenhuis tensor $N_{\phantom{\mu}\nu\rho}^{\mu}$
vanishes identically \cite{Dumitrescu:2012at}, so the almost complex
structure is integrable, and hence $\mathcal{M}$ is a complex manifold.
Furthermore, note that the complex structure is metric-compatible,
i.e. $g_{\mu\nu}J_{\phantom{\mu}\rho}^{\mu}J_{\phantom{\nu}\sigma}^{\nu}=g_{\rho\sigma}$,
so ${\cal M}$ is hermitian.

To simplify some of our later analysis, we introduce holomorphic coordinates
$z^{i},\bar{z}^{i}$ ($i=1,2$), such that 
\begin{equation}
J_{\phantom{i}j}^{i}=i\delta_{j}^{i},\qquad J_{\phantom{i}\bar{j}}^{\bar{i}}=-i\delta_{\bar{j}}^{\bar{i}}.
\end{equation}
One can check that $\Omega_{\bar{i}\bar{j}}=0$, and $\Omega_{12}$
is nonvanishing everywhere. Hence $\Omega$ defines a nowhere vanishing
section of the canonical line bundle ${\cal K}$ of (2,0)-forms. To
summarize, the supersymmetric backgrounds ${\cal M}$ we are considering
are hermitian manifolds with $SU(2)$ structure and trivial canonical
line bundle ${\cal K}$. The only compact 4-manifolds that satisfy
those criteria are tori, $K3$ and primary Kodaira surfaces \cite{Dumitrescu:2012at,Barth:1984}.

\subsection{General invariants}

\label{sub:SU2_gen}In section \ref{sub:triv_gen_inv}, we saw that
imposing constraints on the number of preserved supercharges can lead
to a much richer set of invariants. In this section, we will demonstrate
that the same is true when imposing the condition that the supercharges
are chiral, i.e. for backgrounds with SU(2) structure.

Setting $\epsilon_{R}=0$, the SUSY variations simplify to 
\begin{align}
\delta C & =-\epsilon_{L}^{c}\psi_{L},\label{eq:vec_susy1_SU2}\\
\delta\psi_{L} & =-\epsilon_{L}F,\label{eq:vec_susy2_SU2}\\
\delta\psi_{R} & =\frac{1}{2}\gamma^{\mu}(A_{\mu}+\nabla_{\mu}C)\epsilon_{L},\label{eq:vec_susy3_SU2}\\
\delta F & =0,\label{eq:vec_susy4_SU2}\\
\delta\overline{F} & =\nabla^{\mu}\left(\epsilon_{L}^{c}\gamma_{\mu}\psi_{R}\right)-\epsilon_{L}^{c}\lambda_{L},\label{eq:vec_susy5_SU2}\\
\delta A_{\mu} & =-\epsilon_{L}^{c}\gamma_{\mu}\lambda_{R}+\nabla_{\mu}\left(\epsilon_{L}^{c}\psi_{L}\right),\label{eq:vec_susy6_SU2}\\
\delta\lambda_{L} & =\frac{1}{2}\gamma^{\mu\nu}\epsilon_{L}\nabla_{\mu}A_{\nu}-\frac{1}{2}\epsilon_{L}D,\label{eq:vec_susy7_SU2}\\
\delta\lambda_{R} & =0,\label{eq:vec_susy8_SU2}\\
\delta D & =\nabla^{\mu}\left(\epsilon_{L}^{c}\gamma_{\mu}\lambda_{R}\right)+\frac{2i}{3}b_{\mu}\left(\epsilon_{L}^{c}\gamma^{\mu}\lambda_{R}\right).
\end{align}
The crucial difference to the case of backgrounds with non-chiral supercharges discussed
earlier is that the supercharge $\delta$ is nilpotent: $\delta^{2}=0$.
One fact we can immediately note is that any exact term $\delta V$
will be $\delta$-closed. In particular, localization seems straightforward.
We will further comment on aspects of localization in section \ref{sub:SU2_loc}.

To find all bosonic SUSY invariants, we again make the ansatz 
\begin{equation}
E=\alpha_{1}D+\alpha_{2}F+\alpha_{3}\overline{F}+\alpha_{4}C+\beta^{\mu}A_{\mu},
\end{equation}
with in general nonconstant $\alpha_{i}$ and $\beta^{\mu}$. Demanding
that $\delta E$ is a total derivative, we find the conditions 
\begin{eqnarray}
\nabla^{i}\alpha_{1}-\frac{2}{3}i\alpha_{1}b^{i}+\beta^{i} & = & 0,\\
\alpha_{4}+\nabla^{\mu}\beta_{\mu} & = & 0,\\
\alpha_{3} & = & 0.
\end{eqnarray}
Here $i=1,2$ denote holomorphic coordinates, and we have used the
fact that $\gamma_{\bar{i}}\epsilon_{L}=0$, which follows from Fierz
identities. We do not attempt to find the complete set of solutions,
but instead give three examples of invariants: 
\begin{itemize}
\item From (\ref{eq:vec_susy4_SU2}), we immediately see that $F$-terms
are invariant. Using (\ref{eq:vec_susy2_SU2}), we can show that these
terms are also $\delta$-exact: 
\begin{equation}
\alpha_{2}F=-\delta\left(\alpha_{2}\frac{\epsilon_{L}^{\dagger}\psi_{L}}{f_{L}}\right).\label{eq:Fexact}
\end{equation}
In principle, we can allow $\alpha_{2}$ to be an arbitrary function. 
\item A second type of solution can be obtained by setting $\alpha_{2}=0$,
and restricting $\alpha_{1}$ to be a constant. Since $\alpha_{1}=0$
only leads to a trivial solution, we can set $\alpha_{1}=1$. Then
\begin{eqnarray}
\beta^{i} & = & \frac{2i}{3}b^{i},\nn\\
\alpha_{4} & = & -\nabla^{\mu}\beta_{\mu}.
\end{eqnarray}
There are two linearly independent solutions, characterized by the
choice of $\beta^{\bar{i}}$. We choose the following linearly independent
solutions: 
\begin{eqnarray}
\beta_{1}^{\mu} & = & \frac{2i}{3}b^{\mu},\nn\\
\beta_{2}^{\mu} & = & i\nabla_{\nu}J^{\mu\nu}.
\end{eqnarray}
Note that with this choice, $\beta_{1}^{i}=\beta_{2}^{i}=\frac{2i}{3}b^{i}$
and $\beta_{1}^{\bar{i}}=\frac{2i}{3}b^{\bar{i}}$, but $\beta_{2}^{\bar{i}}=\frac{2i}{3}b^{\bar{i}\star}$.
The corresponding invariants are 
\begin{eqnarray}
E_{1} & = & D+\frac{2i}{3}b^{\mu}A_{\mu}-\frac{1}{3}\left(R+\frac{2}{3}b^{2}\right)C,\nn\\
E_{2} & = & D+i\nabla_{\nu}J^{\mu\nu}A_{\mu}.
\end{eqnarray}
It will be convenient to perform a change of basis by letting 
\begin{equation}
E_{-}\equiv E_{1}-E_{2}=-\mathrm{\frac{4}{3}Im}b^{\overline{i}}A_{\overline{i}}-\frac{1}{2}(R+\frac{2}{3}b^{2})C.
\end{equation}
Using (\ref{eq:vec_susy3_SU2}) and integration by parts, we find
that 
\begin{eqnarray}
E_{-} & = & \delta\left[-\frac{4}{3}\mathrm{Im}b_{\mu}\frac{\epsilon_{L}^{\dagger}\gamma^{\mu}\psi_{R}}{f_{L}}\right],\nn\\
E_{2} & = & \delta\left[-2\frac{\epsilon_{L}^{\dagger}\lambda_{L}}{f_{L}}\right].
\end{eqnarray}

\end{itemize}
We conclude that all three invariants are SUSY-exact, and the partition
function does not depend on the corresponding coupling constants.

In general, $\alpha_{1}$ can be a nontrivial function of the background.
In this more general case, we find that 
\begin{eqnarray}
E & = & \alpha_{1}D+\left(\frac{2i}{3}\alpha_{1}b^{\mu}-\nabla^{\mu}\alpha_{1}\right)A_{\mu}-\nabla^{\mu}\left(\frac{2i}{3}\alpha_{1}b_{\mu}-\nabla_{\mu}\alpha_{1}\right)C\nonumber \\
 & = & \delta\left[-2\alpha_{1}\frac{\epsilon_{L}^{\dagger}\lambda_{L}}{f_{L}}-\frac{4}{3}\left(\alpha_{1}\mathrm{Im}b_{\mu}+\nabla_{\mu}\alpha_{1}\right)\frac{\epsilon_{L}^{\dagger}\gamma^{\mu}\psi_{R}}{f_{L}}\right]+\frac{2}{3}(\Delta_{b}\alpha_{1})C,
\end{eqnarray}
where 
\begin{equation}
\Delta_{b}=-\nabla^{\mu}\nabla_{\mu}-i\nabla_{\mu}J^{\mu\nu}\nabla_{\nu}.
\end{equation}
$E$ is exact if and only if $\Delta_{b}\alpha_{1}=0$, which is clearly
satisfied for $\alpha_{1}=\mathrm{constant}$.

\subsection{Chiral invariants}

\label{sub:SU2_chir}As before, there are additional
chiral and antichiral invariants. These are 
\begin{equation}
F,\quad\overline{F},\quad\overline{\phi},
\end{equation}
evaluated on chiral/antichiral fields. The first two invariants can
be thought of as the special case $M=\overline{M}=0$ of (\ref{eq:IIbar4sc}),
while $\overline{\phi}$ is an additional invariant, due to the form
of the SUSY algebra for SU(2) structure. We find that F-terms are
exact (see (\ref{eq:Fexact})) while $\overline{F}$ and $\overline{\phi}$
are not.

\subsection{Lagrangians and localization}

\label{sub:SU2_loc}It is instructive to compare and contrast the
SU(2) structure case with the case of backgrounds with non-chiral supercharges discussed
in section \ref{sec:trivial}. We will do this by analyzing a simple
toy-model: Consider a pair of chiral and antichiral multiplets $(\Phi,\overline{\Phi})$,
with charges $(1,1$) and $(-1,-1$) under the global $U(1)\times U(1)_{R}$
symmetry. As we saw, on backgrounds with SU(2) structure there is
a bigger arsenal of invariants than for the non-chiral case,
so there is more freedom in building Lagrangians. Supersymmetric Lagrangians
are built by combining superfields into products and taking the corresponding
invariants. For SU(2) structure, there is an additional antisymmetric
product $S_{1}\wedge S_{2}$ (see appendix \ref{sec:tens_calc}),
which gives us even more freedom in constructing Lagrangians. To be
concrete, we can consider the following quadratic Lagrangian:

\begin{eqnarray}
e^{-1}{\cal L} & = & \lambda_{1}\Phi\times\overline{\Phi}\bigg|_{E_{-}}+\lambda_{2}\Phi\times\overline{\Phi}\bigg|_{E_{2}}+\lambda_{3}\Phi\times\overline{\Phi}\bigg|_{F}+\lambda_{4}\Phi\wedge\overline{\Phi}\bigg|_{E_{-}}+\lambda_{5}\Phi\wedge\overline{\Phi}\bigg|_{E_{2}}\nonumber \\
 &  & +\lambda_{F}\Phi\times\Phi\bigg|_{F}+\overline{\lambda}_{F}\overline{\Phi}\times\overline{\Phi}\bigg|_{\overline{F}}.\label{eq:LSU2_sfields}
\end{eqnarray}
The $\lambda_{i}$ are various coupling constants. We have omitted
$\overline{\phi}$-terms, which would break $R$-symmetry explicitly
(see section \ref{sub:SU2_breakR} for a discussion of these terms).

It turns out that not all of the terms in (\ref{eq:LSU2_sfields})
are linearly independent. Using the multiplication rules (\ref{eq:symm_prod})
and (\ref{eq:antisymm_prod}), we can write the Lagrangian in component
form as 
\begin{eqnarray}
e^{-1}{\cal L} & = & t_{1}\delta V_{1}+t_{2}\delta V_{2}+t_{M}\delta V_{M}+t_{b}\delta V_{b}+\lambda_{F}\delta V_{F}\nonumber \\
 &  & +\overline{\lambda}_{F}(2\overline{\phi}\overline{F}+\psi_{R}^{c}\psi_{R}),\label{eq:LSU(2)}
\end{eqnarray}
where 
\begin{align}
V_{1} & =\frac{1}{f_{L}}\epsilon_{L}^{\dagger}\psi_{L}\overline{F},\nn\\
V_{2} & =\frac{1}{f_{L}}\epsilon_{L}^{\dagger}\gamma^{\mu}\psi_{R}\nabla_{\mu}\phi,\nn\\
V_{M} & =-\frac{1}{3f_{L}}\epsilon_{L}^{\dagger}\psi_{L}\overline{\phi},\nn\\
V_{b} & =\frac{2}{3f_{L}}\epsilon_{L}^{\dagger}\gamma^{\mu}\psi_{R}\mathrm{Im}b_{\mu},\nn\\
V_{F} & =-\frac{2}{f_{L}}\epsilon_{L}^{\dagger}\psi_{L}\phi,
\end{align}
and we have chosen a more convenient basis of couplings: 
\begin{eqnarray}
t_{1} & = & -2\lambda_{5},\nn\\
t_{2} & = & 2(\lambda_{2}-\lambda_{5}),\nn\\
t_{M} & = & \frac{3}{2}(\lambda_{2}+\lambda_{3}),\nn\\
t_{b} & = & -2(\lambda_{1}+\lambda_{4}).
\end{eqnarray}
If we set $t_{i}=1$, $\lambda_{F}=-m$ and $\overline{\lambda}_{F}=-\overline{m}$,
the Lagrangian reduces to (\ref{eq:Lfull}), with $K=\overline{\Phi}\Phi$
and $W=m\Phi^{2}$.

The decomposition of (\ref{eq:LSU(2)}) in terms of SUSY-exact terms
makes it manifest that the partition function is independent of all
couplings except $\overline{\lambda}_{F}$. In particular, we are
free to take certain linear combinations of couplings to infinity
to perform localization. We now show that taking $t\equiv t_{1}+t_{2}\rightarrow\infty$
accomplishes just that.

Evaluating the bosonic part of the corresponding ``localization term''
\begin{equation}
t(\delta V_{1}+\delta V_{2})\big|_{\mathrm{bos.}}=t\left(-F\overline{F}+\left(g^{\mu\nu}+iJ^{\mu\nu}\right)\partial_{\mu}\overline{\phi}\partial_{\nu}\phi\right),\label{eq:Locterm_bos}
\end{equation}
we see that it can be made positive semi-definite by choosing the
integration contour $\overline{\Phi}=\Phi^{\ddagger}$ for the bosonic
fields, where $\ddagger$ is the involution 
\begin{equation}
(\phi,\overline{\phi},F,\overline{F})^{\ddagger}=(\overline{\phi},\phi,-\overline{F},-F).\label{eq:realc}
\end{equation}
In the limit $t\rightarrow\infty$, the path integral then localizes
to bosonic field configurations with $(\delta V_{1}+\delta V_{2})\big|_{\mathrm{bos.}}=0$.
In our case, the locus is $F=0$ and $\phi=\phi_{0}=\mathrm{constant}$%
\footnote{A priori, $\phi$ is allowed to be an anti-holomorphic function. However,
on a compact complex manifold this implies that $\phi$ is a constant
\cite{Nakahara:2003} .%
}. The partition function is given by a 1-loop integral around the
classical locus%
\footnote{We neglect the infinite prefactor due to $\int d\phi{}_{0}d\overline{\phi}_{0}$.%
}: 
\begin{equation}
Z=\int{\cal D}\phi{\cal D}\overline{\phi}{\cal D}\psi_{L}{\cal D}\psi_{R}\exp\left[-\int d^{4}x\sqrt{g}\left(\overline{\phi}\Delta_{b}\phi+\psi_{L}^{c}\Delta_{f}\psi_{R}\right)\right].\label{eq:z1loop}
\end{equation}
Here we defined 
\begin{eqnarray}
\mbox{\ensuremath{\Delta}}_{b} & = & -\nabla^{\mu}\nabla_{\mu}-i\nabla_{\mu}J^{\mu\nu}\nabla_{\nu},\nn\\
\Delta_{f} & = & \gamma^{\mu}\left(-\gamma^{5}\nabla_{\mu}+\frac{i}{2}b_{\mu}+\frac{i}{2}\nabla_{\nu}J_{\phantom{\nu}\mu}^{\nu}\right).
\end{eqnarray}

\subsection{Ambiguities of the partition function}

\label{sub:SU2_count} The mere existence of an explicit prescription
(\ref{eq:z1loop}) for calculating the partition function on backgrounds
with SU(2) structure is not sufficient to conclude that $Z$ is a
physical observable. In general, the one-loop determinants that appear
need to be regularized, so it is crucial to ask if the final result
is regularization scheme independent and thus physical. As we saw,
for $SU(2)$ structure the partition function depends nontrivially
only on antichiral couplings $\overline{\lambda}_{F}$. Following
our logic in section \ref{sub:triv_counter}, we should then ask what
possible finite counterterms could render the partition function ambiguous.
The $\overline{F}$-terms in (\ref{eq:LSU2_sfields})
can be viewed as a special case of interactions that arise from 
\begin{equation}
e^{-1}{\cal L}_{\mathrm{int}}=\big[\overline{{\Sigma}}_m \overline{\Phi}^{2}+\overline{\Sigma}_{\lambda}\overline{\Phi}^{3}\big]\big|_{\overline{F}}.
\end{equation}
Here $\overline{\Sigma}_m$ is a spurion that contains a relevant coupling $\overline{m}$ as its lowest component, while $\overline{\Sigma}_{\lambda}$ contains a marginal coupling $\overline{\lambda}$. Since the non-interacting theory is invariant under $U(1)_R$, we can assign R-charges 0 and +1 to $\overline{\Sigma}_m$ and $\overline{\Sigma}_{\lambda}$ to restore R-symmetry.
The only nonzero counterterm consistent with R-symmetry is
\begin{equation}
e^{-1}{\cal L}_{ct}={\overline{\Sigma}}_m^4\big|_{\overline{\phi}},\label{eq:ctm4}
\end{equation}
In particular, there are no mixed matter-curvature counterterms, since the expectation value of the curvature superfield (\ref{eq:Rcurv}) vanishes identically for $SU(2)$ structure, provided that we consider the R-symmetric case $M=0$ (see (\ref{eq:int12})).
We conclude that there is a quartic ambiguity in the free energy:
\begin{equation}
\mathrm{log}Z\sim\mathrm{log}Z+b(\overline{m}r)^4.
\end{equation}
Any terms in $\mathrm{log}Z$ that depend on terms of order $\overline{m}^5$ or higher are free from ambiguities, or in other words,
\begin{equation}
\frac{\partial^5}{\partial(\overline{m}r)^5}\mathrm{log}Z
\end{equation}
is non-ambiguous.

For certain matter contents, additional symmetries may protect
the theory entirely from ambiguities. In fact, this is the case for the toy model
discussed in the previous section. To preserve the global $U(1)$
symmetry, we need to assign nonzero $U(1)$-charges to the spurions. Provided there are no anomalies, the counterterm
(\ref{eq:ctm4}) is simply forbidden,
as it would break $U(1)$. Therefore, in the particular case at
hand, our localization result (\ref{eq:z1loop}) is completely safe from ambiguities,
and thus physical.

More generally, the problem of ambiguities is resolved if we consider
backgrounds that allow for two chiral supercharges, $(\epsilon_{L},0)$
and $(0,\epsilon_{R})$. The problematic $\overline{F}$-terms are
now exact with respect to the additional right-handed supercharge
$\delta_{R}$. As a result, the partition function is completely independent
of couplings, and thus physical. Since the Killing spinor equations
(\ref{eq:KSE12}) are linear and homogeneous, a pair of non vanishing
supercharges $(\epsilon_{L},0)$ and $(0,\epsilon_{R})$ can be combined
into a single supercharge $(\epsilon_{L},\epsilon_{R})$, with the
condition that $M=\overline{M}=0$. Such backgrounds are $T^{2}$-fibrations
over a Riemann surface, which we encountered in section \ref{sec:trivial}.
These backgrounds are therefore ideal candidates to perform localization
(see e.g. \cite{Closset:2013sxa}).

\subsection{Breaking $R$-symmetry}

\label{sub:SU2_breakR}We can explicitly break $R$-symmetry by adding
$\overline{\phi}$-type deformations to our Lagrangian. This corresponds
to adding an antiholomorphic potential
\begin{equation}
{\cal L}\rightarrow{\cal L}+\overline{V}(\overline{\phi}).\label{eq:LWbar}
\end{equation}
In complete analogy to the standard non-renormalization theorems in
flat space \cite{Seiberg:1993vc}, one can show that this does not introduce any additional
finite counterterms involving the couplings $\overline{\lambda}_{\phi}$ within $\overline{V}$. Notice that this result relies crucially on the
fact that even though (\ref{eq:LWbar}) breaks $R$-symmetry, the
background itself is $R$-symmetric. For example, this would not be
the case for theories on $S^{4}$.

Since old-minimal supergravity allows for backgrounds with and without
$R$-symmetry, we can also study the explicit breaking of $U(1)_{R}$
from a supergravity point of view. Looking at (\ref{eq:KSE12}), we
can associate the $M$ and $\overline{M}$-terms with the violation
of $R$-symmetry. In the case of SU(2) structure with supercharge
$(\epsilon_{L},0)$, we have $\overline{M}=0$. The function $M$
however is unconstrained and does not appear in the SUSY variations
or invariants derived above, yet it is still responsible for breaking
R-symmetry: 
Consider the curved superspace interaction 
\begin{equation}
\int d^{2}\overline{\Theta}^{2}2\overline{\epsilon}\overline{W}(\overline{\Phi}).\label{eq:Wbar_sspace}
\end{equation}
For $M\neq0$, the antichiral density is $2\overline{\epsilon}=e\left(1-\overline{\Theta}^{2}M\right)$.
Alternatively, we can recast (\ref{eq:Wbar_sspace}) as a superspace
integral in a background with $M=0$, and treat $2e^{-1}\overline{\epsilon}$
as a spurious antichiral field. Either way, we find 
\begin{equation}
e^{-1}\int d^{2}\overline{\Theta}^{2}2\overline{\epsilon}\overline{W}(\overline{\Phi})=\overline{W}(\overline{\Phi})\big|_{\overline{F}}-M\overline{W}(\overline{\Phi})\big|_{\overline{\phi}}.
\end{equation}
We see that $M$ plays the role of the coupling to the R-violating
$\overline{\phi}$-invariant, which we identify as the antiholomorphic potential $\overline{V}$ in (\ref{eq:LWbar}). It is allowed to be an arbitrary function
because $\delta\overline{\phi}$ vanishes identically, not just up
to total derivatives. Thus turning on a nonzero $M$ corresponds to
breaking $R$-symmetry explicitly.

\section{Discussion}

\label{sec:discussion}In this paper, we have highlighted two unusual
features of $\mathcal{N}=1$ supersymmetry on Euclidean manifolds
with $S^{3}$-isometry (e.g.\ the round and squashed $S^{4}$); namely,
the failure of localization, and regularization scheme dependent ambiguities
of the partition function. Ultimately, both of these features can
be traced back to the structure of off-shell supergravity in the old
minimal formalism. The Killing spinor equation \eqref{eq:KSE12} mixes
left- and right-handed spinors through the $M$- and $\overline{M}$-terms.
This has the consequence that there are backgrounds that admit only non-chiral
Killing spinors of the form $(\epsilon_{L},\epsilon_{R})$, where
the left- and right-handed components cannot be ``disentangled''.
This is manifest in the fact that the supercharge squares to a complex
generator $\delta^{2}\sim{\cal L}_{K}$, with $K^{\mu}=\epsilon_{R}^{c}\gamma^{\mu}\epsilon_{L}$
being a complex Killing vector that mixes left and right chiralities.
Since $\delta$ does not square to an obvious symmetry of the theory,
it appears that SUSY-exact terms are in general not SUSY-closed. In
this paper, we have proven an equivalent statement, namely that there
are no supersymmetric invariants (SUSY-closed terms) that can be written as SUSY-exact
terms. We have explicitly identified the obstruction to exactness
in terms of the non-vanishing Killing vector $L=[K,K^{\star}]$, which
generates part of the isometry group $SU(2)\times SU(2)$ of $S^{3}$.

While the above obstruction might not appear to be very deep at first,
it has the important consequence that the partition function must
depend nontrivially on the values of all coupling constants. We have
discussed two important corollaries: First, adding any term to the Lagrangian will necessarily change the theory, 
so the partition function cannot be calculated using localization. A crucial point in arriving at this result was the classification
of general SUSY-invariants, which generically only need to preserve a single supercharge. Since we showed
that none of these invariants is exact, there are simply no allowed deformations of the theory, and the partition function
does not localize. 

Second, we have shown that there are finite supergravity counterterms that
introduce scheme-dependent ambiguities into the partition function. Our results extend beyond the previously studied
case of SCFTs on $S^{4}$ \cite{Gerchkovitz:2014gta} to any four-dimensional supersymmetric background
with $S^{3}$-isometry. 
While in the conformal case it was shown that the finite part of the partition function is completely unphysical, our analysis demonstrates that $\mathrm{log}Z$ depends on relevant couplings in such a way that ambiguities are under control: If we expand the free energy in powers of relevant couplings, we find
\begin{equation}
\mathrm{log}Z(m,\lambda)=\mathrm{log}Z(0,\lambda)+\sum_{i=1}^{3}(mr)^{i}a_{i}(\lambda)+\tilde{F}(mr,\lambda),
\end{equation}
where the $a_i$ are functions of the marginal couplings, and $\tilde{F}$ may contain all powers of $mr$ except $n=0,1,2,3$ . On $S^4$, the $\mathrm{log}Z(0,\lambda)$-term can be interpreted as the free energy of the CFT, which is subject to ambiguities, and thus unphysical. As we have shown, the terms up to cubic order in $m$ are ambiguous as well. However, the higher-order part $\tilde{F}$ is free from ambiguities and thus physical. 

A similar feature has been observed for $\mathcal{N}=2^\star$ theories on $S^4$, where the partition function can been computed using either localization \cite{Pestun:2007rz} or holographic techniques \cite{Bobev:2013cja}. 
It would be interesting to calculate the unambiguous part $\tilde{F}$ of the free energy for the $\mathcal{N}=1$ case as well, and explicitly confirm some of the results of this paper.


An obvious way to avoid the complications present in the $S^{3}$-isometry
case is to consider only backgrounds for which the chirality-mixing
terms in \eqref{eq:KSE12} vanish identically. This has led us to
analyze backgrounds with $U(1)_{R}$ $R$-symmetry, which possess
at least one nilpotent supercharge, $\delta^{2}=0$. In this case,
many simplifications occur: With one exception (anti-chiral $\overline{F}$-terms),
the partition function does not depend on the values of couplings
in our Lagrangian \eqref{eq:LSU(2)}, and localization is straightforward.
However, the fact that we have found a procedure for calculating the
partition function does not necessarily mean that the result will
be sensible. As we demonstrated in section \ref{sub:SU2_count}, 
the partition function is in general subject to antiholomorphic ambiguities. Interestingly, the only ambiguity appears at quartic order in relevant couplings, and thus renormalizes the cosmological constant. This is a special feature of the BRST-like symmetry $\delta$, which provides a trivial extension of the isometry algebra of the background. Some of the standard arguments in Lorentzian supersymmetry, such as the proof of non-renormalization of the vacuum energy, therefore do not apply.

Finally, for backgrounds that preserve two supercharges
of opposite chirality, $Z$ is completely independent of all couplings, and there are no ambiguities. Within the framework of old-minimal supergravity,
the only manifolds with this property are torus-fibrations over two-dimensional Riemann surfaces. It would be interesting to
carry out localization for explicit cases of such backgrounds, presumably paralleling
the analysis in \cite{Closset:2013sxa,Nishioka:2014zpa}.

There are two caveats to our analysis of ambiguities of partition
functions in sections \ref{sub:triv_counter} and \ref{sub:SU2_count}, which point towards interesting future directions:
First, our classification of possible finite counterterms necessarily
requires the existence of a regularization scheme that preserves the symmetries of the theory. As far as we know, there
is not yet a satisfactory answer to the question when such a scheme
does or does not exist for a supersymmetric theory. If for a given
theory there is no supersymmetric regularization scheme, conclusions
about the partition function, such as independence of couplings and
the physical content, would need to be reexamined. Second,
we have only analyzed finite counterterms that involve both matter couplings and curvature invariants at the same time.
It would be interesting to also analyze purely gravitational counterterms, which arise as
$F$-type and $D$-type terms evaluated on the various
curvature multiplets of supergravity \cite{Assel:2014tba}.


\section*{Acknowledgments}

{We are grateful to H. Elvang for insightful discussions on counterterm
ambiguities in supergravity and helpful comments on a first draft of this paper. We also thank C. Closset, A. Faraggi,
V. Pestun and Y. Ruan for useful discussions.} This work is supported
in part by the US Department of Energy under grant DE-SC0007859.


\appendix

\section{Supersymmetric tensor calculus}

\label{sec:tens_calc}In order to construct supersymmetric Lagrangians,
we need to know the rules for combining superfields \cite{Ferrara:1978jt,Ferrara:1978wj,Stelle:1978yr}.
Given two multiplets $S_{1}$ and $S_{2}$, we can form a new multiplet
\begin{eqnarray}
S_{1}\times S_{2} & \equiv & \left(C_{12},\psi_{12L},\psi_{12R},F_{12},\overline{F}_{12},A_{12\mu},\lambda_{12L},\lambda_{12R},D_{12}\right).
\end{eqnarray}
Demanding that $C_{12}=C_{1}C_{2}$, we can work out the multiplication
rules using (\ref{eq:vec_susy19}): 
\begin{align}
C_{12} & =C_{1}C_{2},\nonumber \\
\psi_{12L} & =C_{1}\psi_{2L}+C_{2}\psi_{1L},\nonumber \\
\psi_{12R} & =C_{1}\psi_{2R}+C_{2}\psi_{1R},\nonumber \\
F_{12} & =C_{1}F_{2}+C_{2}F_{1}-\psi_{1L}^{c}\psi_{2L},\nonumber \\
\overline{F}_{12} & =C_{1}\overline{F}_{2}+C_{2}\overline{F}_{1}+\psi_{1R}^{c}\psi_{2R},\nonumber \\
A_{12\mu} & =C_{1}A_{2\mu}+\psi_{1L}^{c}\gamma_{\mu}\psi_{2R}+(1\leftrightarrow2),\nonumber \\
\lambda_{12L} & =C_{1}\lambda_{2L}+\overline{F}_{1}\psi_{2L}-\frac{1}{2}\gamma^{\mu}(A_{1\mu}-\nabla_{\mu}C_{1})\psi_{2R}+(1\leftrightarrow2),\nonumber \\
\lambda_{12R} & =C_{1}\lambda_{2R}+F_{1}\psi_{2R}+\frac{1}{2}\gamma^{\mu}(A_{1\mu}+\nabla_{\mu}C_{1})\psi_{2L}+(1\leftrightarrow2),\nonumber \\
D_{12} & =C_{1}D_{2}+2F_{1}\overline{F}_{2}+2\psi_{1R}^{c}\lambda_{2R}-2\psi_{1L}^{c}\lambda_{2L}+\psi_{1L}^{c}\gamma^{\mu}(\nabla_{\mu}-\frac{i}{2}b_{\mu})\psi_{2R}\nonumber \\
 & \quad-\psi_{1R}^{c}\gamma^{\mu}(\nabla_{\mu}+\frac{i}{2}b_{\mu})\psi_{2L}+\frac{1}{2}(A_{1}^{\mu}A_{2\mu}-\nabla^{\mu}C_{1}\nabla_{\mu}C_{2})+(1\leftrightarrow2).\label{eq:symm_prod}
\end{align}
It is easy to see that the product operator $\times$ is symmetric,
i.e. $S_{1}\times S_{2}=S_{2}\times S_{1}$. This is a result of demanding
$C_{12}=C_{1}C_{2}$. A natural question is whether there also exists
an antisymmetric product $\wedge$, such that $C_{12}=0$. We can
attempt to derive the multiplication rules in a similar fashion, starting
with 
\begin{equation}
0=\delta C_{12}=-\epsilon_{L}^{c}\psi_{12L}-\epsilon_{R}^{c}\psi_{12R}.\label{eq:deltaC_anti}
\end{equation}
A quick check reveals that for the case of non-chiral supercharges, all the
components of the product multiplet have to be set to zero, i.e. there
is no nontrivial antisymmetric product. In the chiral (or SU(2) structure) case,
however, we have more freedom: Setting $\epsilon_{R}=0$, we see that
Eqn.~(\ref{eq:deltaC_anti}) is solved by $\psi_{12L}=0$, but nonzero
$\psi_{12R}$. In fact, we find that there exists an antisymmetric
product 
\begin{equation}
S_{1}\wedge S_{2}\equiv\left(C_{12},\psi_{12L},\psi_{12R},F_{12},\overline{F}_{12},A_{12\mu},\lambda_{12L},\lambda_{12R},D_{12}\right),
\end{equation}
with the following multiplication rules: 
\begin{align}
C_{12} & =0,\nonumber \\
\psi_{12L} & =0,\nonumber \\
\psi_{12R} & =C_{1}\psi_{2R}-C_{2}\psi_{1R},\nonumber \\
F_{12} & =0,\nonumber \\
\overline{F}_{12} & =C_{1}\overline{F}_{2}-C_{2}\overline{F}_{1},\nonumber \\
A_{12\mu} & =C_{1}(A_{2\mu}+\nabla_{\mu}C_{2})+\psi_{1L}^{c}\gamma_{\mu}\psi_{2R}-(1\leftrightarrow2),\nonumber \\
\lambda_{12L} & =C_{1}\lambda_{2L}-\overline{F}_{1}\psi_{2L}+\gamma^{\mu}\nabla_{\mu}C_{1}\psi_{2R}-(1\leftrightarrow2),\nonumber \\
\lambda_{12R} & =C_{1}\lambda_{2R}+F_{1}\psi_{2R}-\frac{1}{2}\gamma^{\mu}(A_{1\mu}+\nabla_{\mu}C_{1})\psi_{2L}-(1\leftrightarrow2),\nonumber \\
D_{12} & =C_{1}D_{2}+2F_{1}\overline{F}_{2}-2\psi_{1L}^{c}\lambda_{2L}+\psi_{1L}^{c}\gamma^{\mu}(\nabla_{\mu}-\frac{i}{2}b_{\mu})\psi_{2R}\nonumber \\
 & +\psi_{1R}^{c}\gamma^{\mu}(\nabla_{\mu}+\frac{i}{2}b_{\mu})\psi_{2L}+A_{1}^{\mu}\nabla_{\mu}C_{2}-(1\leftrightarrow2).\label{eq:antisymm_prod}
\end{align}
Similar expressions can be derived for the case $\epsilon_{L}=0$.

\bibliography{SusyFT-bib}

\providecommand{\href}[2]{#2}\begingroup\raggedright\begin{thebibliography}{10}

\bibitem{Pestun:2007rz}
V.~Pestun, {\it {Localization of gauge theory on a four-sphere and
  supersymmetric Wilson loops}},  \href{http://xxx.lanl.gov/abs/0712.2824}{{\tt
  arXiv:0712.2824}}.

\bibitem{Drukker:2000rr}
N.~Drukker and D.~J. Gross, {\it {An exact prediction of N = 4 SUSYM theory for
  string theory}},  {\em J. Math. Phys.} {\bf 42} (2001) 2896--2914,
  [\href{http://xxx.lanl.gov/abs/hep-th/0010274}{{\tt hep-th/0010274}}].

\bibitem{Erickson:2000af}
J.~K. Erickson, G.~W. Semenoff, and K.~Zarembo, {\it {Wilson loops in N = 4
  supersymmetric Yang-Mills theory}},  {\em Nucl. Phys.} {\bf B582} (2000)
  155--175, [\href{http://xxx.lanl.gov/abs/hep-th/0003055}{{\tt
  hep-th/0003055}}].

\bibitem{Kapustin:2009kz}
A.~Kapustin, B.~Willett, and I.~Yaakov, {\it {Exact Results for Wilson Loops in
  Superconformal Chern- Simons Theories with Matter}},  {\em JHEP} {\bf 03}
  (2010) 089, [\href{http://xxx.lanl.gov/abs/0909.4559}{{\tt
  arXiv:0909.4559}}].

\bibitem{Marino:2011nm}
M.~Marino, {\it {Lectures on localization and matrix models in supersymmetric
  Chern-Simons-matter theories}},  {\em J.Phys.} {\bf A44} (2011) 463001,
  [\href{http://xxx.lanl.gov/abs/1104.0783}{{\tt arXiv:1104.0783}}].

\bibitem{Festuccia:2011ws}
G.~Festuccia and N.~Seiberg, {\it {Rigid Supersymmetric Theories in Curved
  Superspace}},  \href{http://xxx.lanl.gov/abs/1105.0689}{{\tt
  arXiv:1105.0689}}.

\bibitem{Jia:2011hw}
B.~Jia and E.~Sharpe, {\it {Rigidly Supersymmetric Gauge Theories on Curved
  Superspace}},  {\em JHEP} {\bf 1204} (2012) 139,
  [\href{http://xxx.lanl.gov/abs/1109.5421}{{\tt arXiv:1109.5421}}].

\bibitem{Samtleben:2012gy}
H.~Samtleben and D.~Tsimpis, {\it {Rigid supersymmetric theories in 4d
  Riemannian space}},  {\em JHEP} {\bf 1205} (2012) 132,
  [\href{http://xxx.lanl.gov/abs/1203.3420}{{\tt arXiv:1203.3420}}].

\bibitem{Klare:2012gn}
C.~Klare, A.~Tomasiello, and A.~Zaffaroni, {\it {Supersymmetry on Curved Spaces
  and Holography}},  \href{http://xxx.lanl.gov/abs/1205.1062}{{\tt
  arXiv:1205.1062}}.

\bibitem{Dumitrescu:2012ha}
T.~T. Dumitrescu, G.~Festuccia, and N.~Seiberg, {\it {Exploring Curved
  Superspace}},  \href{http://xxx.lanl.gov/abs/1205.1115}{{\tt
  arXiv:1205.1115}}.

\bibitem{Liu:2012bi}
J.~T. Liu, L.~A. Pando~Zayas, and D.~Reichmann, {\it {Rigid Supersymmetric
  Backgrounds of Minimal Off-Shell Supergravity}},  {\em JHEP} {\bf 1210}
  (2012) 034, [\href{http://xxx.lanl.gov/abs/1207.2785}{{\tt
  arXiv:1207.2785}}].

\bibitem{deMedeiros:2012sb}
P.~de~Medeiros, {\it {Rigid supersymmetry, conformal coupling and twistor
  spinors}},  {\em JHEP} {\bf 1409} (2014) 032,
  [\href{http://xxx.lanl.gov/abs/1209.4043}{{\tt arXiv:1209.4043}}].

\bibitem{Dumitrescu:2012at}
T.~T. Dumitrescu and G.~Festuccia, {\it {Exploring Curved Superspace (II)}},
  {\em JHEP} {\bf 1301} (2013) 072,
  [\href{http://xxx.lanl.gov/abs/1209.5408}{{\tt arXiv:1209.5408}}].

\bibitem{Sohnius:1981tp}
M.~F. Sohnius and P.~C. West, {\it {An Alternative Minimal Off-Shell Version of
  N=1 Supergravity}},  {\em Phys.Lett.} {\bf B105} (1981) 353.

\bibitem{Closset:2013vra}
C.~Closset, T.~T. Dumitrescu, G.~Festuccia, and Z.~Komargodski, {\it {The
  Geometry of Supersymmetric Partition Functions}},  {\em JHEP} {\bf 1401}
  (2014) 124, [\href{http://xxx.lanl.gov/abs/1309.5876}{{\tt
  arXiv:1309.5876}}].

\bibitem{Closset:2014uda}
C.~Closset, T.~T. Dumitrescu, G.~Festuccia, and Z.~Komargodski, {\it {From
  Rigid Supersymmetry to Twisted Holomorphic Theories}},  {\em Phys.Rev.} {\bf
  D90} (2014) 085006, [\href{http://xxx.lanl.gov/abs/1407.2598}{{\tt
  arXiv:1407.2598}}].

\bibitem{Ferrara:1978jt}
S.~Ferrara and P.~van Nieuwenhuizen, {\it {Tensor Calculus for Supergravity}},
  {\em Phys.Lett.} {\bf B76} (1978) 404.

\bibitem{Ferrara:1978wj}
S.~Ferrara and P.~Van~Nieuwenhuizen, {\it {Structure of Supergravity}},  {\em
  Phys.Lett.} {\bf B78} (1978) 573.

\bibitem{Stelle:1978yr}
K.~Stelle and P.~C. West, {\it {Tensor Calculus for the Vector Multiplet
  Coupled to Supergravity}},  {\em Phys.Lett.} {\bf B77} (1978) 376.

\bibitem{Wess:1992cp}
J.~Wess and J.~Bagger, {\em {Supersymmetry and supergravity}}.
\newblock Univ. Pr., Princeton, USA, 1992.

\bibitem{Stelle:1978wj}
K.~Stelle and P.~C. West, {\it {Relation Between Vector and Scalar Multiplets
  and Gauge Invariance in Supergravity}},  {\em Nucl.Phys.} {\bf B145} (1978)
  175.

\bibitem{Zamolodchikov:1986gt}
A.~Zamolodchikov, {\it {Irreversibility of the Flux of the Renormalization
  Group in a 2D Field Theory}},  {\em JETP Lett.} {\bf 43} (1986) 730--732.

\bibitem{Kutasov:1988xb}
D.~Kutasov, {\it {Geometry on the Space of Conformal Field Theories and Contact
  Terms}},  {\em Phys.Lett.} {\bf B220} (1989) 153.

\bibitem{Gerchkovitz:2014gta}
E.~Gerchkovitz, J.~Gomis, and Z.~Komargodski, {\it {Sphere Partition Functions
  and the Zamolodchikov Metric}},  {\em JHEP} {\bf 1411} (2014) 001,
  [\href{http://xxx.lanl.gov/abs/1405.7271}{{\tt arXiv:1405.7271}}].

\bibitem{Jockers:2012dk}
H.~Jockers, V.~Kumar, J.~M. Lapan, D.~R. Morrison, and M.~Romo, {\it
  {Two-Sphere Partition Functions and Gromov-Witten Invariants}},  {\em
  Commun.Math.Phys.} {\bf 325} (2014) 1139--1170,
  [\href{http://xxx.lanl.gov/abs/1208.6244}{{\tt arXiv:1208.6244}}].

\bibitem{Gomis:2012wy}
J.~Gomis and S.~Lee, {\it {Exact Kahler Potential from Gauge Theory and Mirror
  Symmetry}},  {\em JHEP} {\bf 1304} (2013) 019,
  [\href{http://xxx.lanl.gov/abs/1210.6022}{{\tt arXiv:1210.6022}}].

\bibitem{Doroud:2012xw}
N.~Doroud, J.~Gomis, B.~Le~Floch, and S.~Lee, {\it {Exact Results in D=2
  Supersymmetric Gauge Theories}},  {\em JHEP} {\bf 1305} (2013) 093,
  [\href{http://xxx.lanl.gov/abs/1206.2606}{{\tt arXiv:1206.2606}}].

\bibitem{Benini:2012ui}
F.~Benini and S.~Cremonesi, {\it {Partition functions of N=(2,2) gauge theories
  on $S^2$ and vortices}},  \href{http://xxx.lanl.gov/abs/1206.2356}{{\tt
  arXiv:1206.2356}}.

\bibitem{Doroud:2013pka}
N.~Doroud and J.~Gomis, {\it {Gauge theory dynamics and Kähler potential for
  Calabi-Yau complex moduli}},  {\em JHEP} {\bf 1312} (2013) 99,
  [\href{http://xxx.lanl.gov/abs/1309.2305}{{\tt arXiv:1309.2305}}].

\bibitem{Assel:2014tba}
B.~Assel, D.~Cassani, and D.~Martelli, {\it {Supersymmetric counterterms from
  new minimal supergravity}},  \href{http://xxx.lanl.gov/abs/1410.6487}{{\tt
  arXiv:1410.6487}}.

\bibitem{Stelle:1978ye}
K.~Stelle and P.~C. West, {\it {Minimal Auxiliary Fields for Supergravity}},
  {\em Phys.Lett.} {\bf B74} (1978) 330.

\bibitem{Ferrara:1978em}
S.~Ferrara and P.~van Nieuwenhuizen, {\it {The Auxiliary Fields of
  Supergravity}},  {\em Phys.Lett.} {\bf B74} (1978) 333.

\bibitem{Sohnius:1982fw}
M.~Sohnius and P.~C. West, {\it {The Tensor Calculus and Matter Coupling of the
  Alternative Minimal Auxiliary Field Formulation of N=1 Supergravity}},  {\em
  Nucl.Phys.} {\bf B198} (1982) 493.

\bibitem{Nosaka:2013cpa}
T.~Nosaka and S.~Terashima, {\it {Supersymmetric Gauge Theories on a Squashed
  Four-Sphere}},  {\em JHEP} {\bf 1312} (2013) 001,
  [\href{http://xxx.lanl.gov/abs/1310.5939}{{\tt arXiv:1310.5939}}].

\bibitem{Lust:2010by}
D.~Lust, P.~Patalong, and D.~Tsimpis, {\it {Generalized geometry, calibrations
  and supersymmetry in diverse dimensions}},  {\em JHEP} {\bf 1101} (2011) 063,
  [\href{http://xxx.lanl.gov/abs/1010.5789}{{\tt arXiv:1010.5789}}].

\bibitem{Barth:1984}
W.~Barth, C.~Peters, and A.~Van~de Ven, {\em Compact Complex Surfaces}.
\newblock Springer, 1984.

\bibitem{Nakahara:2003}
M.~Nakahara, {\em {Geometry, topology and physics}}.
\newblock Taylor \& Francis, 2003.

\bibitem{Closset:2013sxa}
C.~Closset and I.~Shamir, {\it {The $\mathcal{N}=1$ Chiral Multiplet on
  $T^2\times S^2$ and Supersymmetric Localization}},  {\em JHEP} {\bf 1403}
  (2014) 040, [\href{http://xxx.lanl.gov/abs/1311.2430}{{\tt
  arXiv:1311.2430}}].

\bibitem{Seiberg:1993vc}
N.~Seiberg, {\it {Naturalness versus supersymmetric nonrenormalization
  theorems}},  {\em Phys.Lett.} {\bf B318} (1993) 469--475,
  [\href{http://xxx.lanl.gov/abs/hep-ph/9309335}{{\tt hep-ph/9309335}}].

\bibitem{Bobev:2013cja}
N.~Bobev, H.~Elvang, D.~Z. Freedman, and S.~S. Pufu, {\it {Holography for $N =
  2^*$ on $S^4$}},  {\em JHEP} {\bf 1407} (2014) 001,
  [\href{http://xxx.lanl.gov/abs/1311.1508}{{\tt arXiv:1311.1508}}].

\bibitem{Nishioka:2014zpa}
T.~Nishioka and I.~Yaakov, {\it {Generalized indices for $ \mathcal{N} $ = 1
  theories in four-dimensions}},  {\em JHEP} {\bf 1412} (2014) 150,
  [\href{http://xxx.lanl.gov/abs/1407.8520}{{\tt arXiv:1407.8520}}].

\end{thebibliography}\endgroup
{} \bibliographystyle{JHEP}
\end{document}